\documentclass[reprint,aps,prb,amsmath,floatfix,amssymb]{revtex4-2}
\usepackage{mathtools, bm, xcolor, graphicx, hyperref}

\usepackage{multirow}

\usepackage[capitalise]{cleveref}

\graphicspath{ {./figs/} }

\begin{document}

\title{
Multiple insulating states due to the interplay of strong correlations and lattice geometry in a single-orbital Hubbard model
}
\author{H. L. Nourse}
\affiliation{School of Mathematics and Physics, The University of Queensland, Brisbane, Queensland 4072, Australia}
\author{Ross H. McKenzie}
\affiliation{School of Mathematics and Physics, The University of Queensland, Brisbane, Queensland 4072, Australia}
\author{B. J. Powell}
\affiliation{School of Mathematics and Physics, The University of Queensland, Brisbane, Queensland 4072, Australia}

\begin{abstract}
We report ten ground states arising from strong correlations in the single-orbital Hubbard model on the decorated honeycomb lattice; including Dirac metals, flat-band ferromagnets, real-space Mott insulators, dimer and trimer Mott insulators, and a spin-$1$ Mott insulator. The rich phase diagram arises from structures within the unit cell. Hence, such states are absent on simpler lattices. We argue that such insulating phases are prevalent on decorated lattices. These are found in many materials and common in coordination polymers, providing a playground to explore this physics.
\end{abstract}

\maketitle

Decorated lattices are found in a wide range of materials, including inorganic compounds \cite{Zheng2007,Bao2015,Nguyen2018,Taniguchi1995,Ye2011,Bao2011}, organometallics \cite{Jacko2015}, organic molecular crystals \cite{Shuku2018}, and coordination polymers (CPs) \cite{Murase2017,Murase2017b,Kingsbury2017,Jeon2015,Darago2015,DeGayner2017,Henling2014,Henline2014,Polunin2015,Kalmutzki2018}. They consist of one or more cluster types, e.g., a molecule, linked to form a net \cite{Wells1977,Kalmutzki2018}. Many materials with decorated lattices are reported to have novel ground states \cite{Ueda1996,Janani2014b,Nourse2016,Reja2019,Ruegg2010,Wen2010,Chen2018,Lopez2019,Yao2007,Sur2018,Dagotto2013,Yanagi2014,Khatami2014,Feng2019}. Perhaps the simplest is the decorated honeycomb lattice (DHL), realized in materials such as the trinuclear organometallic  compounds, e.g., Mo$_3$S$_7$(dmit)$_3$ \cite{Jacko2015}, in organic molecular crystals \cite{Shuku2018}, in iron (III) acetates \cite{Zheng2007}, in cold fermionic atoms \cite{Lin2014}, and in CPs \cite{Henling2014,Henline2014,Polunin2015}. There are a number of theoretical studies that predict exotic phases of matter on this lattice, such as the quantum spin Hall insulator \cite{Ruegg2010}, quantum anomalous Hall insulator \cite{Wen2010,Chen2018,Lopez2019}, topological metals \cite{Lopez2019}, valence bond solids (VBS) \cite{Richter2004,Misguich2007,Yang2010,Jahromi2018}, and quantum spin liquids \cite{Kitaev2006,Yao2007b,Dusuel2008,Khosla2017,Powell2017} with non-Abelian anyons \cite{Kitaev2006,Yao2007b,Dusuel2008}. 

Rich phase diagrams often arise from the complex interplay of interactions between multiple orbitals, as found in the discovery and analysis of the superconducting pnictide compounds \cite{Si2016}. Here, we report a rich phase diagram with only a single orbital and an on-site Hubbard repulsion, but multiple sites in the unit cell. This suggests an alternative minimal path to rich physics arising from the unique structure of decorated lattices that is not found in simpler lattices.

\begin{figure}
	\centering
	\includegraphics[width=\columnwidth]{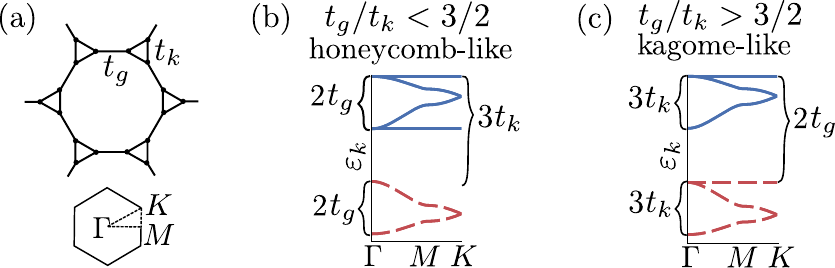}
	\caption{\label{fig:lattice}
		(a) A triangle decorates each vertex of the honeycomb lattice.
		(b) Non-interacting band structure of the DHL with strong intra-triangle hopping. The dashed (solid) line has mostly $A$ ($E$) trimer orbital character (cf. \cref{fig:clusters}e). (c) Strong inter-triangle hopping. The dashed (solid) line has mostly (anti-)bonding orbital character (cf. \cref{fig:clusters}b).
	}
\end{figure}

\begin{figure}
	\centering
	\includegraphics[width=\columnwidth]{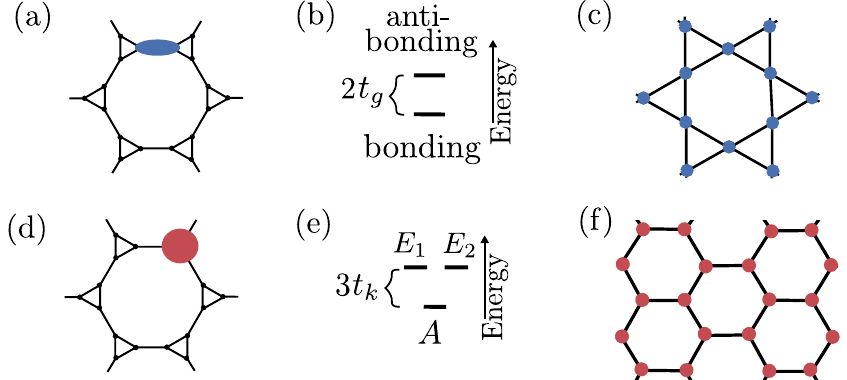}
	\caption{\label{fig:clusters}
		Clusters used.
		(a) Clustering the DHL as dimers maps to a (b) two-orbital model on the (c) kagome lattice.
		(d) Clustering the lattice as trimers maps to a (e) three-orbital model on the (f) honeycomb lattice.
		In both cases there are complicated inter- and intra-orbital hopping between clusters with phases that depend on direction.
	}
\end{figure}

In this Letter we study the single-orbital Hubbard model on the DHL (\cref{fig:lattice}a). Despite the model having only three parameters ($t_g/t_k$, $U/t_k$, $n$), we find a plethora of interaction driven phases (\cref{tab:summary-phases}). Some of the insulators occur away from half-filling where a metal is expected. These arise because of effective multi-orbital interactions due to the structures that decorate the lattice. We construct simple pictures of these insulating phases by studying appropriate `molecular' limits -- analogous to the atomic limit for the usual Mott-Hubbard transition. Low-energy effective theories of these insulators include the spin-$1/2$ Heisenberg model on the kagome lattice, and the spin-$1/2$ and spin-$1$ Heisenberg models on the honeycomb lattice (\cref{fig:lattice,fig:clusters}). We argue that such `molecular' Mott insulators are prevalent to decorated lattices. With the chemical flexibility found in CPs, decorated lattices provides an avenue to explore rich physics in condensed matter systems.

\begin{table} \centering %\small %\ra{1.2}
	\begin{tabular}{p{0.2\columnwidth} p{0.6\columnwidth} p{0.2\columnwidth}}
		%\toprule
		%
		\multirow{1}{4em}{\centering $n$} & \hfil $t_g \lesssim t_k$ & \multirow{1}{4em}{\centering $\mathcal{S}$} \\
		\colrule
		\multirow{2}{4em}{\centering 1/3}  & \hfil Trimer Mott & \multirow{2}{4em}{\centering 1/2} \\
		[-1.1ex] & \hfil (Honeycomb N\'{e}el order)  & \\
		\multirow{1}{4em}{\centering 2/3}  & \hfil Band insulator & \\
		\multirow{1}{4em}{\centering 5/6}  & \hfil Flat-band ferromagnet & \\
		\multirow{2}{4em}{\centering 1}  & \hfil Real-space Mott & \multirow{2}{4em}{\centering 1/2} \\ 
		[-1.1ex] & \hfil (Broken $\mathcal{C}_3$ VBS) & \\
		\multirow{2}{4em}{\centering 4/3}  & \hfil Spin-1 Mott & \multirow{2}{4em}{\centering 1} \\
		[-1.1ex] & \hfil (Honeycomb N\'{e}el order) & \\
		\multirow{1}{4em}{\centering 11/6} & \hfil Flat-band ferromagnet & \\ \\
		\multirow{1}{4em}{\centering $n$} & \hfil $t_g \gtrsim t_k$ & \multirow{1}{4em}{\centering $\mathcal{S}$} \\
		\colrule
		\multirow{2}{4em}{\centering 1/2}  & \hfil Dimer Mott & \multirow{2}{4em}{\centering 1/2} \\
		[-1.1ex] & \hfil (Kagome QSL?) & \\
		\multirow{1}{4em}{\centering 5/6} &  \hfil Flat-band ferromagnet & \\
		\multirow{2}{4em}{\centering 1}  & \hfil $t_g$-dimer VBS  (large $U$) crossover & 
		\multirow{2}{4em}{\centering 1/2} \\ 
		[-1.1ex] & \hfil  to band insulator (small $U$) & \\
		\multirow{2}{4em}{\centering 3/2} & \hfil Dimer Mott & \multirow{2}{4em}{\centering 1/2}  \\
		[-1.1ex] & \hfil (Kagome QSL?) & \\
		\multirow{1}{4em}{\centering 11/6} & \hfil Flat-band ferromagnet \\
		%
	%	\botrule
	\end{tabular}
	\caption[Summary of the correlated insulating phases of matter.]
	{\label{tab:summary-phases}
		Summary of the insulating states of the single-orbital Hubbard model on the decorated honeycomb lattice (DHL), where $t_g$ ($t_k$) are the inter-(intra-)triangle hopping amplitudes (\cref{fig:lattice}a), and $n$ is the filling per site. The ground state candidates of the effective spin-$\mathcal{S}$ Heisenberg model in the Mott insulating phases are in parentheses. QSL denotes quantum spin liquid and VBS denotes valence bond solid.
	}
\end{table}

The Hamiltonian for the Hubbard model on the DHL \cite{Jacko2015} is
\begin{align} \label{eq:hubb}
\hat{H} \equiv & - t_g \sum_{\langle i \alpha, j \alpha \rangle, \sigma} \hat{c}_{i\alpha\sigma}^{\dagger} \hat{c}_{j\alpha\sigma}^{}
- t_k \sum_{i,\alpha\neq\beta ,\sigma} \hat{c}_{i\alpha\sigma}^{\dagger} \hat{c}_{i\beta\sigma}^{} \nonumber \\
& + U\sum_{i,\alpha} \hat{n}_{i\alpha\uparrow} \hat{n}_{i\alpha\downarrow},
\end{align}
where $\hat{c}_{i\alpha\sigma}^{\dagger}$ ($\hat{c}_{i\alpha\sigma}^{}$) creates (annihilates) an electron with spin $\sigma \in \{\uparrow,\downarrow\}$ on site $\alpha \in \{1,2,3\}$ of triangle cluster $i$, $\hat{n}_{i\alpha\sigma} \equiv \hat{c}_{i\alpha\sigma}^{\dagger} \hat{c}_{i\alpha\sigma}^{}$, $t_g$ ($t_k$) is the inter-(intra-)triangle hopping integral (\cref{fig:lattice}a), $\langle i\alpha, j\alpha\rangle$ signifies nearest-neighbor hopping between sites of adjacent triangles, and $U$ is the local Coulomb repulsion. 

We use rotationally invariant slave bosons (RISB) \cite{Kotliar1986,Lechermann2007,Lanata2015,Lanata2017} to find the ground state of \cref{eq:hubb} using two-site (\cref{fig:clusters}a-c) or three-site clusters (\cref{fig:clusters}d-f), which are the minimal cluster sizes to capture the intra-cluster correlations necessary to describe the Mott insulators that we find away from half-filling in the kagome-like regime ($t_g/t_k > 3/2$) and honeycomb-like regime ($t_g/t_k < 3/2$) respectively. Technical details can be found in the Supplemental Material \cite{sup}. Equivalent to the Gutzwiller approximation for multi-orbital models \cite{Bunemann2007}, RISB renormalizes an uncorrelated wavefunction by adjusting the weights of local electronic configurations on a cluster, and at the mean-field level the metallic state is a simple realization of a Landau Fermi liquid. RISB  successfully describes many strongly correlated phenomena \cite{Burdin2000,Zhu2004,Lechermann2009,Lu2013,Lanata2015,Lanata2017,Isidori2019}. The use of RISB allows us to capture important spatial correlations necessary for the insulators that we find, it describes exactly the limits of isolated trimers ($t_g = 0)$ and isolated dimers ($t_k = 0$), and allows us to study a wide range of parameters at a reasonable computational cost.

\emph{Insulators in RISB.}--Only the coherent part of the many-body Green's function matrix $\bm{G}(k,\omega)$ is captured within RISB. An effect of the correlations in the metallic state is narrowing of the quasiparticle bands $\xi_p(k)$ (a band is indexed by $p$), and a loss of spectral weight, captured by the quasiparticle weight $Z_p^{\mathrm{qp}}(k)$ \cite{sup}. For quasiparticle bands that cross the Fermi energy ($\omega = 0$) RISB describes a Fermi liquid where $Z_p^{\mathrm{qp}}(k)$ is a measure of the metallicity, with $Z_p^{\mathrm{qp}}(k) = 1$ corresponding to a non-interacting metal and $Z_p^{\mathrm{qp}}(k) = 0$ to a correlated insulator, where the Fermi surface vanishes. This is the usual description of a Mott insulator at half-filling captured in slave boson theories \cite{Kotliar1986}, and originally described by Brinkman and Rice \cite{Brinkman1970}.

We work in the basis of the molecular orbitals (\cref{fig:clusters}) where the quasiparticle weight is a diagonal matrix $\bm{Z}$ and assume that the clusters are uniform. For two-site clusters $\bm{Z}_i = \mathrm{diag} (Z_{b}, Z_{a})$ and for three-site clusters $\bm{Z}_i = \mathrm{diag} (Z_A, Z_{E_1}, Z_{E_2})$ with $Z_E \equiv Z_{E_1} = Z_{E_2}$. In the honeycomb-like regime (\cref{fig:lattice}b) the lower (upper) bands have mostly $A$ ($E$) orbital character, and in the kagome-like regime (\cref{fig:lattice}c) the lower (upper) bands have mostly (anti-)bonding orbital character. To a good approximation the quasiparticle weight in the molecular orbital basis approximates the quasiparticle weight in the bands -- they are exactly equivalent at the $\Gamma$ and $K$ points \cite{sup}.

\begin{figure}
	\centering
	\includegraphics[width=\columnwidth]{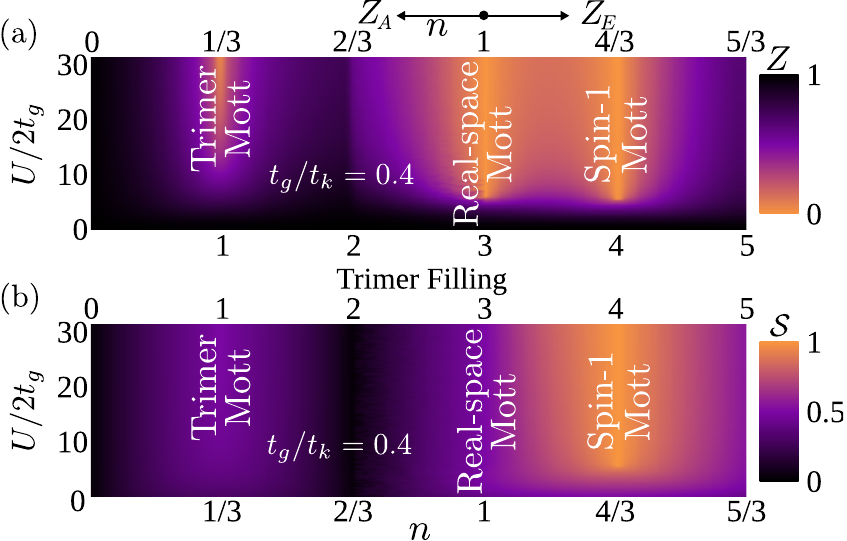} % final
	\caption{\label{fig:heatmaps-trimer}
		Phase diagram in the honeycomb-like regime for the triangle clusters (\cref{fig:clusters}d-f).
		(a) The relevant matrix element of $\bm{Z}$ that approximates the quasiparticle weight for bands at the Fermi energy. The quasiparticle weight vanishes in the Mott insulating phases. For $n \le 1$ ($n>1$) we show the quasiparticle weight $Z_A$ ($Z_E$) associated with the $A$ ($E$) trimer orbital. At $n=1$ $Z_A \sim Z_E$.
		(b) Effective spin $\mathcal{S}$ of a triangle, where $\mathcal{S}$ is the solution to $\mathcal{S}(\mathcal{S} + 1) = \sum_{i} \langle \vec{S}_i \cdot \vec{S}_i \rangle / 2 \mathcal{N}$, the spin of triangle $i$ is $\vec{S}_{i} = \sum_{\alpha=1}^3 \frac{1}{2} \sum_{\sigma \sigma'} \hat{c}_{i\alpha\sigma}^{\dagger} \vec{\tau}_{\sigma\sigma'} \hat{c}_{i\alpha\sigma'}^{}$, $\vec{\tau}$ is a vector of Pauli matrices, and $\mathcal{N}$ is the number of unit cells. 
		A spin-$1/2$ degree of freedom arises on each triangle in the real-space and trimer Mott insulators. A spin-$1$ moment occurs in the spin-$1$ Mott insulator because an effective intra-triangle Hund's coupling $\tilde{J}$ favors the formation of spin-triplets on a triangle \cite{Merino2006}.
	}
\end{figure}

\begin{figure}
	\centering
	\includegraphics[width=1\columnwidth]{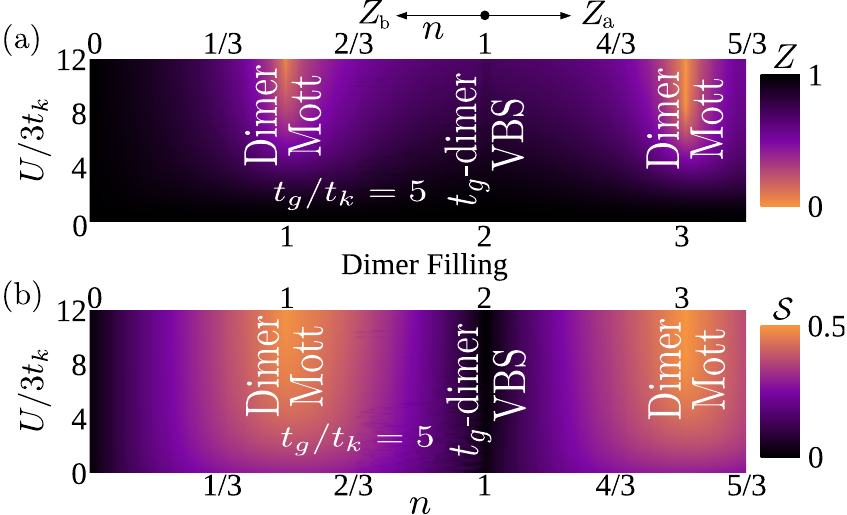} % final
	\caption{\label{fig:heatmaps-dimer}
		Phase diagram in the kagome-like regime for the dimer clusters (\cref{fig:clusters}a-c).
		(a) The relevant matrix element of $\bm{Z}$ that approximates the quasiparticle weight for bands at the Fermi energy. The quasiparticle weight vanishes in the Mott insulating phases. For $n \le 1$ ($n>1$) we show the quasiparticle weight $Z_b$ ($Z_a$) associated with the (anti-)bonding dimer orbital. At $n=1$ $Z_b \sim Z_a$.
		(b) Effective spin $\mathcal{S}$ of a dimer, where $\mathcal{S}(\mathcal{S} + 1) = \sum_{i} \langle \vec{S}_i \cdot \vec{S}_i \rangle / 3 \mathcal{N}$, and the spin of dimer $i$ is $\vec{S}_{i} = \sum_{\alpha=1}^2 \frac{1}{2} \sum_{\sigma \sigma'} \hat{c}_{i\alpha\sigma}^{\dagger} \vec{\tau}_{\sigma\sigma'} \hat{c}_{i\alpha\sigma'}^{}$.
		A spin-$1/2$ degree of freedom arises on each dimer in the dimer Mott insulators. At half-filling spin-singlet formation along the $t_g$ bond leads to $\mathcal{S} = 0$.
	}
\end{figure}

\emph{Real-space Mott insulator.}--The usual place to look for a Mott insulator is at half-filling ($n=1$) because $U$ disfavors double occupancy and suppresses charge fluctuations so that electrons become localized to a lattice site. For $t_g/t_k \le 3/2$ and at finite $U$ a metal-insulator transition occurs where the quasiparticle weight for all bands vanishes (\cref{fig:heatmaps-trimer}a). For $t_g / t_k > (t_g / t_k)_c = 0.9-1.0$ the dominating electronic configurations are spin-singlets along the $t_g$ bonds (\cref{fig:heatmaps-dimer}b), while for $t_g / t_k < (t_g / t_k)_c$ spin-singlets form along the $t_k$ bonds. In the latter phase, there is an instability to a magnetic state that breaks the $\mathcal{C}_3$ rotational symmetry of the lattice \cite{sup}.

The Heisenberg model on the DHL with exchange coupling $J_g = 4t_g^2 / U$ and $J_k = 4 t_k^2 / U$ provides an effective model of the Hubbard model at  large-$U$ and $n=1$. This model has been extensively studied  \cite{Richter2004,Misguich2007,Yang2010,Jahromi2018}. Based on a detailed numerical study it has been argued that for $J_g / J_k \gtrsim 0.9$ ($t_g / t_k \gtrsim 0.95$) a $J_g$-dimer VBS forms, while for $J_g / J_k \lesssim 0.9$ there is a VBS that favors spin-singlet formation along $J_k$ bonds but breaks the $\mathcal{C}_3$ rotational symmetry of a triangle \cite{Jahromi2018}. That our RISB calculations, which deal with the full fermionic degrees of freedom, qualitatively agrees with calculations for the spin model (and gives a good estimate for the critical coupling) suggests that RISB is a good approximation for the problem at hand.

Away from half-filling the naive expectation is a correlated metal because on average there is not one electron per site. However, structure of the unit cell of decorated lattices provides important additional degrees of freedom that are not present in simpler lattices. Below we will show that these structures lead to several different interaction driven insulators (\cref{fig:heatmaps-trimer,fig:heatmaps-dimer}), as is the case for multi-orbital atomic systems.

\emph{Dimer Mott insulators.}--When inter-triangle hopping is strong ($t_g/t_k > 3/2$) significant insight can be gained from working in the orbital basis of a dimer formed along the $t_g$ bonds (\cref{fig:clusters}a-c). The Hamiltonian describing a dimer has bonding and anti-bonding orbitals separated in energy by $2t_g$, with intra-orbital and inter-orbital Coulomb repulsion ($\tilde{U} = U/2$), and inter-orbital spin-flip terms \cite{sup}. This leads to a model analogous to the Hubbard bilayer model \cite{Lechermann2007} on the kagome lattice with complicated inter-site hopping that mixes the bonding and anti-bonding orbitals. 

In the dimer molecular limit ($t_g, U  \rightarrow \infty$ with $U/t_g$ finite) RISB with the cluster shown in Fig. \ref{fig:clusters}b becomes exact.
The mixing between the bonding and anti-bonding orbitals vanishes and the electronic structure contains two decoupled copies of the kagome lattice (\cref{fig:lattice}c) with an intra-orbital repulsion $\tilde{U}$. Half-filling the (anti-)bonding orbital and turning on interactions is formally equivalent to the half-filled single-orbital Hubbard model on the kagome lattice \cite{sup}. A Mott insulator occurs for sufficiently large $\tilde{U}$ \cite{Ohashi2006} and in the limit $\tilde{U} \rightarrow \infty$ is connected to the ground state of isolated dimers. 

Our RISB results  show that the dimer Mott insulators are extended phases away from these limits. In \cref{fig:heatmaps-dimer}a we show the quasiparticle weight $Z_b$ ($Z_a$) of the (anti-)bonding orbitals for $n \le 1$ ($>1$), which correspond to the lower (upper) bands in the kagome-like regime (\cref{fig:lattice}c). An insulator occurs at (three-)quarter-filling for sufficiently large $U$ with the bonding orbitals half-filled (fully occupied) and the anti-bonding orbitals empty (half-filled). Interactions  renormalize the electrons so that the dimer orbitals decouple, and the insulator is adiabatically connected to the insulator in the dimer molecular  limit: a Mott insulator on the kagome lattice. As $t_g/t_k$ is reduced the dimer Mott insulator becomes unstable and there is instead a metal, occurring for $t_g/t_k \lesssim 4.1$ $(3.0)$ for $n=1/2$ ($3/2$).

The low-energy physics of the dimer Mott insulators is crucially different to the Mott insulator at half-filling. At half-filling the low-energy effective theory is the spin-$1/2$ Heisenberg model on the DHL. In the dimer Mott insulators charge fluctuations are suppressed between dimers with each dimer forming a spin $\mathcal{S} = 1/2$ moment (\cref{fig:heatmaps-dimer}b). Hence,  the low-energy effective theory of the dimer Mott insulators is the spin-$1/2$ Heisenberg model on the kagome lattice, whose ground state may be a quantum spin liquid \cite{Broholm2020}.

These dimer Mott insulators on the DHL are similar to those observed in the organic charge transfer salts $\kappa$-(BEDT-TTF)$_2 X$, where the BEDT-TTF molecules form a dimer and share one hole \cite{Kanoda1997,McKenzie1998,Powell2006}. For many $X$ the intra-dimer hopping is more than twice the inter-dimer hopping \cite{Jacko2019} and a minimal model to describe the insulator is the half-filled single-orbital Hubbard model on the anisotropic triangular lattice \cite{McKenzie1998,Powell2006}. Our results demonstrate that such insulators also arise in decorated lattices.

\emph{Trimer Mott insulator.}--The simplest way to understand the insulating phases in the limit of strong intra-triangle hopping ($t_g/t_k < 3/2$)  is  to work in the eigenbasis of a triangle formed by the $t_k$ bonds (\cref{fig:clusters}d-f). The Hamiltonian describing a trimer has an $A$ orbital separated in energy from two degenerate $E$ orbitals by $3t_k$, with intra-orbital and inter-orbital Coulomb repulsion $\tilde{U} = U/3$, a Hund's coupling $\tilde{J} = -\tilde{U}/3$ that favors spin-triplet formation on a triangle \cite{Merino2006,Janani2014a,Janani2014b,Nourse2016}, inter-orbital spin-flip, and correlated inter-orbital hopping terms \cite{sup}. The Hamiltonian becomes a three-orbital model on the honeycomb lattice with complicated inter-site hopping.

In the trimer molecular limit ($t_k, U \rightarrow\infty$, $U/t_k$ finite) RISB with the cluster shown in Fig. \ref{fig:clusters}d is exact. At one-sixth-filling ($n=1/3$)  the ground state has one electron in the $A$ orbital with empty $E$ orbitals. In this limit the Hamiltonian is formally equivalent to the single-orbital Hubbard model on the honeycomb lattice \cite{sup}. A Mott insulator occurs for large $\tilde{U}$ and at $\tilde{U} \rightarrow \infty$ is connected to the ground state of isolated trimers. 

RISB demonstrates that the trimer Mott insulator is stable away from the trimer molecular limit, \cref{fig:heatmaps-trimer}a. However, for sufficiently small $\tilde{U}$ charge fluctuations between the $A$ and $E$ orbitals destroy the insulating state and a Dirac metal is recovered. The trimer Mott insulator is adiabatically connected to the Mott insulator of the half-filled single-orbital Hubbard model on the honeycomb lattice. Hence, the low-energy effective theory is the spin-$1/2$ Heisenberg model on the honeycomb lattice whose ground state is N\'{e}el ordered \cite{Mattsson1994,Banerjee2011,Pujari2013,Block2013,Harada2013,Gong2013,Yu2014,Bishop2015}. Above a critical hopping ratio $t_g/t_k \gtrsim 0.45$ the trimer Mott insulator is not found because a finite $U$ is not sufficient to decouple the $A$ and $E$ orbitals and localize electrons.

\emph{Spin-$1$ Mott insulator.}--At two-thirds filling ($n=2/3$)  the trimer $E$ orbitals are degenerate and the effective intra-orbital interactions become crucial for understanding the insulating state. In the trimer molecular limit RISB is exact and, for $U > 0$, the ground state  contains a spin-triplet on each trimer due to the effective Hund's interaction $\tilde{J}$.
For $t_k\gg t_g$ and $n>1/3$ the Hamiltonian maps to a  two-orbital Hubbard-Kanamori model on the honeycomb lattice with  inter-site hopping intergrals whose phases are direction dependent \cite{sup}. 

In \cref{fig:heatmaps-trimer}a we show the quasiparticle weight $Z_E$ of the $E$ orbitals for $n > 1$, which corresponds to the upper bands in the honeycomb-like regime (\cref{fig:lattice}b). We find that $Z_E$ vanishes, resulting in a metal-insulator transition from a Dirac metal to a spin-$1$ Mott insulator with the $E$ ($A$) orbitals half-filled (fully occupied). In the insulator the degenerate $E$ orbitals are decoupled from the $A$ orbitals with each triangle forming a spin-triplet because of the effective Hund's interaction $\tilde{J}$ (\cref{fig:heatmaps-trimer}b). Within RISB the spin-$1$ Mott insulator is adiabatically connected to the isolated trimer limit and a Mott insulator on the effective two-orbital model at half-filling on the honeycomb-like lattice. Above the critical hopping parameter ratio $t_g / t_k \gtrsim 0.86$ no spin-$1$ Mott insulator exists for finite $U$.

The low energy effective theory of the spin-$1$ Mott insulator is the spin-$1$ Heisenberg model \cite{Merino2016,Powell2017} on the honeycomb lattice whose ground state is N\'{e}el ordered \cite{Merino2018,Merino2018b,Gong2015,Campbell2016}. The molecular crystal Mo$_3$S$_7$(dmit)$_3$ is two-thirds filled and an isolated monolayer is the DHL \cite{Jacko2015}. Hence, we propose that isolated monolayers of Mo$_3$S$_7$(dmit)$_3$ are spin-$1$ N\'{e}el ordered.

Crucially, the spin-$1$ Mott insulator requires the effective Hund's coupling $\tilde{J}$ on a triangle. This can straightforwardly be confirmed by varying the effective multi-orbital interactions. The dimer and trimer Mott insulators occur even when there is only an effective intra-orbital Coulomb repulsion $\tilde{U}$ with no multi-orbital interactions. In contrast, there is no Mott insulating phase at two-thirds filling with only intra-orbital $\tilde{U}$. A metal-insulator transition only occurs when the multi-orbital interactions on a trimer are included. 

\emph{Ferromagnetism.}--Mielke and Tasaki proved that the ground state of the Hubbard model with a flat band is a ferromagnetic insulator for $U>0$ provided criteria are satisfied \cite{Mielke1991,Mielke1992,Tasaki1992}, which our model meets for any $t_g/t_k$ and when the upper flat band is half-filled ($n = 11/6$). The criteria for the rigorous proof are not met by the lower flat band. Nevertheless,  RISB predicts ferromagnetic long-range order when the lower flat band is partially filled ($2/3 < n < 1$) \cite{sup}.

Additionally, the DHL has van Hove singularities at fillings $n=1/4, 5/12, 5/4$, and $17/2$, where $\rho(E) \rightarrow \infty$. Due to the Stoner mechanism \cite{Stoner1951} the ground state is a ferromagnetic metal near $n=1/4$ and $n=5/12$. However, antiferromagnetic correlations dominate near $n=5/4$ and $n=17/2$ because of the proximity to the spin-$1$ Mott insulator \cite{sup}.

\emph{Conclusions.}--Based on the above results, we propose that molecular Mott insulators are general on decorated lattices. The bases that diagonalize  local structures within the unit cell of decorated lattices provide an intuitive picture of the insulating phases that are stabilized by strong electronic correlations. If the local structure is sufficiently complicated then correlations can drive more exotic ground states, as demonstrated by the spin-$1$ Mott insulator at two-thirds filling on the DHL. 

Many materials realize the DHL lattice \cite{Jacko2015,Shuku2018,Zheng2007,Lin2014,Henling2014,Henline2014,Polunin2015}. Electronic structure calculations place some in the kagome-like \cite{Henline2014,Shuku2018} and others in the  honeycomb-like \cite{Polunin2015,Jacko2015} regimes. Several of these materials have insulating phases that have not previously had a detailed theoretical explanation. Chemical doping could allow for the exploration of their phase diagrams. More broadly, decorated lattices are  common in CPs and found in many other classes of materials. These are typically insulating -- but detailed theoretical explanations of these insulating states are largely absent. Our work provides the theoretical framework for understanding these materials.

An important open question is: do unconventional superconducting states generically arise near these molecular Mott insulating phases? The superconductivity in the dimer Mott insulators in $\kappa$-(BEDT-TTF)$_2X$ \cite{Powell2006,Powell2005,Powell2007} and some multiorbital models  \cite{Hoshino2015,Hoshino2016,Reja2019} and materials  \cite{Si2016} suggests that it may.

\emph{Note added in proof}. Near half-filling there is singlet superconductivity with extended-s, extended-d, and f-wave symmetry \cite{Merino2020}.

\begin{acknowledgments}
	We thank Jason Pillay and Elise Kenny for helpful conversations. This work was supported by the Australian Research Council through Grants No. DP160102425, DP160100060, and DP181006201.
\end{acknowledgments}

%\bibliography{bibliography}

%

%%%%%%%%%% Merge with supplemental materials %%%%%%%%%%
%\pagebreak
\clearpage
%\widetext
\onecolumngrid
\begin{center}
\textbf{\large Supplemental Material: Multiple insulating states due to the interplay of strong correlations and lattice geometry in a single-orbital Hubbard model}
\end{center}
%%%%%%%%%% Merge with supplemental materials %%%%%%%%%%
%%%%%%%%%% Prefix a "S" to all equations, figures, tables and reset the counter %%%%%%%%%%
\setcounter{equation}{0}
\setcounter{figure}{0}
\setcounter{table}{0}
\setcounter{section}{0}
\setcounter{page}{1}
%\makeatletter
\renewcommand{\theequation}{S\arabic{equation}}
\renewcommand{\thefigure}{S\arabic{figure}}
\renewcommand{\bibnumfmt}[1]{[S#1]}
\renewcommand{\citenumfont}[1]{S#1}
\renewcommand{\thetable}{S\arabic{table}}
\renewcommand{\thesection}{S\Roman{section}}
%%%%%%%%%% Prefix a "S" to all equations, figures, tables and reset the counter %%%%%%%%%%

\vspace{0.3cm}
\twocolumngrid

\section{Rotationally invariant slave bosons} \label{sec:supp-RISB}

The results in the main text on the decorated honeycomb lattice (DHL) are obtained from the mean-field approximation to rotationally invariant slave bosons (RISB) \cite{SLechermann2007, SIsidori2009, SLanata2015, SLanata2017}. The approach is a good description of Landau Fermi liquid behavior and metal-insulator transitions in multi-orbital systems at a moderate computational cost. Within RISB, a simpler description is obtained by mapping the system onto an auxiliary fermionic Hamiltonian which is purely quadratic, with renormalized hopping due to local interactions described by the bosons of the theory, which are treated at the saddle-point level. In this way, RISB captures the coherent low-energy quasiparticles. High-energy incoherent excitations are associated with fluctuations around the saddle-point, which are not considered in this work.

Consider a generic electronic Hamiltonian given by
\begin{equation} \label{eq:supp-ham}
	\hat{H} \equiv \hat{H}^{\mathrm{kin}} + \sum_i \hat{H}_i^{\mathrm{loc}},
\end{equation}
with
\begin{align}
	\hat{H}^{\mathrm{kin}} \equiv - \sum_{i\neq j \alpha \beta} [t_{ij}]_{\alpha \beta} \hat{c}_{i\alpha}^{\dagger} \hat{c}_{j\beta}^{},
\end{align}
\begin{equation}
	\hat{H}_i^{\mathrm{loc}} = \sum_{AB} \langle A_i | \hat{H}_i^{\mathrm{loc}} | B_i \rangle |A_i \rangle \langle B_i |
\end{equation}
where $[t_{ij}]_{\alpha \beta}$ is the hopping amplitude between clusters $i$,$j$, $\alpha = 1, \ldots, M_i$ labels all electronic flavors within a cluster (sites, orbitals, and spin), and the local Fock states of cluster $i$ are given by 
\begin{equation}
	|A_i\rangle = [\hat{c}_{i1}^{\dagger}]^{\nu_1} \ldots [\hat{c}_{iM_i}^{\dagger}]^{\nu_{M_i}} |0\rangle,  \quad (\nu_{\alpha} = 0,1),
\end{equation}
where $A_i = 1\ldots 2^{M_i}$ runs over all the possible lists of occupation numbers $\{ \nu_1 , \ldots , \nu_{M_i} \}$, and $|0\rangle$ is the vacuum. To explicitly connect with the DHL given by Eq.~1 in the main text for triangular clusters (Fig.~2d-f of the main text), the $t_g$ term is the kinetic part between clusters, while the $t_k$ and $U$ terms are the local part within a cluster. In simple slave boson theories \cite{SKotliar1986} only local density-density interactions have a faithful representation. For clusters of the DHL even though the $U$ term is a density-density interaction, the $t_g$ term will hybridize the orbitals and introduce off-diagonal correlations between sites within a cluster. Therefore, a rotationally invariant formalism is required to treat the intra-cluster correlations correctly and capture the insulators presented in the main text.

In RISB theory, first presented in \cite{SLechermann2007}, a new set of auxiliary fermions are introduced for each electronic flavor within a cluster $\{ \hat{f}_{ia} | a = 1, \ldots, M_i \}$ with Fock states $|n_i \rangle$. For each pair of physical and auxliary fermion states $(|A_i\rangle, |n_i\rangle)$ a boson is introduced $\hat{\Phi}_{iAn}$, with the restriction that the auxiliary fermion and physical electron states have the same number of particles $N_{A} = N_{n}$ (for superconductivity where this restriction must be relaxed see \cite{SIsidori2009}). Therefore, the local physical Hilbert space of electronic states of a cluster is mapped onto a larger Hilbert space, with the physical states in the enlarged space given by
\begin{equation}
	|\underline{A}_i \rangle \equiv \frac{1}{\sqrt{D_A}} \sum_n \hat{\Phi}_{iAn}^{\dagger} |0 \rangle \otimes |n_i\rangle = \mathcal{U} |A_i\rangle,
\end{equation}
where $D_A = {M_i \choose N_A}$ is the number of auxiliary fermion states with particle number equal to $N_A$, and $\mathcal{U}$ is the unitary operator that defines the mapping between the original physical Hilbert space and the physical states within the enlarged Hilbert space. The computational complexity grows with the size of the enlarged Hilbert space as $2^{2M_i}$.

The physical electron operator must be defined to act as
\begin{equation} \label{eq:supp-matrix-elements}
	\underline{\hat{c}}_{i\alpha}^{\dagger} |\underline{A}_i\rangle = \sum_{B} \langle B_i | \hat{c}_{i\alpha}^{\dagger} | A_i \rangle |\underline{B}_i \rangle,
\end{equation}
so that the matrix elements of $\underline{\hat{c}}_{i\alpha}^{\dagger}$ in the enlarged Hilbert space are the same as the matrix elements of $\hat{c}_{i\alpha}^{\dagger}$ in the original Hilbert space of the physical electron. A faithful representation is given by
\begin{equation}
	\underline{\hat{c}}_{i\alpha}^{\dagger} \equiv \sum_{a} \hat{\mathcal{R}}_{ia\alpha} \hat{f}_{ia}^{\dagger},
\end{equation}
where
\begin{equation}
	\hat{\mathcal{R}}_{ia\alpha} \equiv \sum_{AB} \sum_{nm} \frac{\langle A_i | \hat{c}_{i\alpha}^{\dagger}|B_i\rangle \langle n| \hat{f}_{ia}^{\dagger} | m \rangle}{\sqrt{N_A(M_i - N_B)}} \hat{\Phi}_{iAn}^{\dagger} \hat{\Phi}_{iBm}^{},
\end{equation}
which becomes the renormalization matrix $\mathcal{R}_i$ of the inter-cluster hopping parameters at mean-field.

To restrict the enlarged Hilbert space to only the subset of physical states the following equations are enforced (the Gutzwiller constraints),
\begin{align}
	\hat{K}_{i}^0 & \equiv \sum_{An}  \hat{\Phi}_{iAn} \hat{\Phi}_{iAn} - \hat{1}, \label{eq:supp-const1} \\
	\hat{K}_{iab} & \equiv \hat{f}_{ia}^{\dagger} \hat{f}_{ib}^{} - \sum_{Anm} \langle m_i| \hat{f}_{ia}^{\dagger} \hat{f}_{ib}^{} | n_i \rangle \hat{\Phi}_{iAn} \hat{\Phi}_{iAm}, \label{eq:supp-const2}
\end{align}
where $\hat{1}$ is the identity.

Any local observable acting on a cluster can be written as
\begin{equation}
	\hat{X} [\underline{\hat{c}}_{i\alpha}^{\dagger}, \underline{\hat{c}}_{i\alpha}^{} ] = \sum_{AB} \langle A_i| \hat{X} [\hat{c}_{i\alpha}^{\dagger}, \hat{c}_{i\alpha}^{}] | B_i \rangle \sum_n \hat{\Phi}_{iAn}^{\dagger} \hat{\Phi}_{iBn}^{}.
\end{equation}
Hence, the representation of the Hamiltonian (\cref{eq:supp-ham}) in the enlarged Hilbert space is \emph{exactly} given by
\begin{equation}
	\underline{\hat{H}} \equiv \underline{\hat{H}}^{\mathrm{kin}} + \sum_i \underline{\hat{H}}_i^{\mathrm{loc}} + \sum_i \underline{\hat{H}}_i^{\mathrm{const}},
\end{equation}
with
\begin{align}
	\underline{\hat{H}}^{\mathrm{kin}} &= \sum_{i\neq j,\alpha\beta} t_{ij}^{\alpha\beta} \underline{\hat{c}}_{i\alpha}^{\dagger} \underline{\hat{c}}_{i\beta}^{}, \\
	\underline{\hat{H}}_i^{\mathrm{loc}} & \equiv \sum_{AB} \langle A_i | \hat{H}_i^{\mathrm{loc}} | B_i \rangle \sum_n \hat{\Phi}_{iAn}^{\dagger} \hat{\Phi}_{iBn}^{}, \\
	\underline{\hat{H}}_i^{\mathrm{const}} & \equiv E_i^c \hat{K}_i^0 + \sum_{ab} \lambda_{iab} \hat{K}_{iab},
\end{align}
where $E_i^c$ and $[\lambda_i]_{ab}$ are Lagrange multiplier fields to enforce the constraints \cref{eq:supp-const1,eq:supp-const2}.

In this representation the fermions only enter the problem quadratically and can be integrated out, leaving a purely bosonic action that depends on the slave bosons and Lagrange multiplier fields.
The mean-field saddle-point approximation is obtained by the replacement of the bosonic fields with their average values, 
$ \Phi_{iAn}(\tau) \rightarrow [\phi_i]_{An} $,
$ E_i^c(\tau) \rightarrow E_i^c $
$ \lambda_{iab}(\tau) \rightarrow [\lambda_i]_{ab} $, which have to be obtained self-consistently by extremizing the free-energy, which at zero temperature is given by
\begin{align} \label{eq:supp-free-energy}
	\Omega & [\phi, E^c, \Lambda]  = \lim_{\beta \rightarrow \infty} \frac{1}{\beta} \sum_{i\omega_n} \mathrm{Tr} \ln (i\omega_n - \hat{H}^{\mathrm{qp}} )^{-1} \nonumber \\
	& + \sum_{iABnm} [\phi_i^{\dagger}]_{nA} \left[ \langle A_i| \hat{H}_i^{\mathrm{loc}} |B_i\rangle \delta_{nm} - E^c \delta_{AB} \delta_{nm} \phantom{\sum_{ab}} \right. \nonumber \\
	& + \left. \sum_{ab} [\lambda_i]_{ab} \langle n_i| \hat{f}_{ia}^{\dagger} \hat{f}_{ib} |m_i \rangle \right] [\phi_i^{}]_{Bm} - E_i^c,
\end{align}
where $i\omega_n$ is a Matsubara frequency, $\beta$ is the inverse temperature, and the Hamiltonian of the quasiparticles is given by
\begin{align} \label{eq:ham-qp}
	\hat{H}^{\mathrm{qp}} & \equiv - \sum_{ij,\alpha\beta,ab} [t_{ij}]_{\alpha \beta} [\mathcal{R}_i^{}]_{a\alpha} [\mathcal{R}_{j}^{\dagger}]_{\beta b} \hat{f}_{ia}^{\dagger} \hat{f}_{jb}^{} \nonumber \\
	& \phantom{{} \equiv} + \sum_{i,ab} [\lambda_i]_{ab} \hat{f}_{ia}^{\dagger} \hat{f}_{ib}^{}.
\end{align}
It is common to interpret the single-particle spectrum of $\hat{H}^{\mathrm{qp}}$ as an approximation to the true coherent spectrum of quasiparticles of the system (\cref{eq:supp-spectral-weight}) \cite{SBunemann2003,SLanata2017b}.

Solutions to \cref{eq:supp-free-energy} are usually obtained in the thermodynamic limit and the clusters are assumed to be homogeneous so that Bloch's theorem applies. In $k$-space of the unit-cell (or super-cell) the renormalization matrices $\bm{\mathcal{R}}$ and correlation potential matrices $\bm{\lambda}$ are block diagonal, with each block indexed by $\ell$ and labeling an inequivalent cluster within the unit cell. For example, on three-site clusters of the DHL the two triangular clusters $\ell = \mathcal{A}, \mathcal{B}$ within the unit-cell are inequivalent and $\bm{\mathcal{R}} \equiv \mathrm{diag}(\bm{\mathcal{R}}_{\mathcal{A}}, \bm{\mathcal{R}}_{\mathcal{B}})$, $\bm{\lambda} \equiv \mathrm{diag}(\bm{\lambda}_{\mathcal{A}}, \bm{\lambda}_{\mathcal{B}})$. On two-site clusters of the DHL there are three inequivalent dimers within a unit-cell. 

The physical electron Green's function is a matrix given by
\begin{align}
	\bm{G}(k,\omega)^{-1} & = \bm{\mathcal{R}}^{-1} \bm{G}^{\mathrm{qp}}(k,\omega)^{-1} \bm{\mathcal{R}}^{\dagger -1} \nonumber \\
	& = \omega \left( \bm{\mathcal{R}}^{\dagger} \bm{\mathcal{R}}^{} \right)^{-1} - \bm{\varepsilon}(k) - \bm{\mathcal{R}}^{-1} \bm{\lambda} \bm{\mathcal{R}}^{\dagger -1} ,
\end{align}
where $\bm{\varepsilon}(k)$ is the dispersion relation between clusters given by $\hat{H}^{\mathrm{kin}}$, $\bm{G}^{\mathrm{qp}}(k,\omega) \equiv (\omega \bm{1} - \bm{\mathcal{H}}^{\mathrm{qp}}(k))^{-1}$ is the Green's function of the auxiliary fermions with
\begin{equation}
	\bm{\mathcal{H}}^{\mathrm{qp}}(k) \equiv \bm{\mathcal{R}}^{} \bm{\varepsilon}(k) \bm{\mathcal{R}}^{\dagger} + \bm{\lambda},
\end{equation}
and $\bm{1}$ is the identity. Comparing to the non-interacting Green's function of the physical electrons, given by
\begin{equation}
	\bm{G}^{0}(k,\omega)^{-1} = \omega \bm{1} - \bm{\varepsilon}(k) - \bm{\varepsilon}^0 ,
\end{equation}
the self-energy is identified as
\begin{align}
	\bm{\Sigma}(\omega) & \equiv \bm{G}^0(k,\omega)^{-1} - \bm{G}_{}(k,\omega)^{-1}  \nonumber \\
	& = \omega\left( \bm{1} - \left( \bm{\mathcal{R}}^{\dagger} \bm{\mathcal{R}}^{} \right)^{-1} \right) + \bm{\mathcal{R}}^{-1} \bm{\lambda} \bm{\mathcal{R}}^{\dagger -1} - \bm{\varepsilon}^0,
\end{align}
where $\bm{\varepsilon}^0$ are the one-body terms present in $\hat{H}^{\mathrm{loc}}$. Note that the self-energy is entirely local and block diagonal in the clusters. Within RISB only the coherent part of $\bm{G}(k,\omega)$ is captured and describes the quasiparticles.

The spectral function matrix $\bm{A}(k,\omega) \equiv -\pi^{-1} \mathrm{Im} \bm{G}(k,\omega)$ within RISB is given by
\begin{equation} \label{eq:supp-spectral-weight}
	\bm{A}(k,\omega) = \bm{\mathcal{R}}^{\dagger} \delta\left(\omega \bm{1} - \bm{\mathcal{H}}^{\mathrm{qp}}(k) \right) \bm{\mathcal{R}}^{}. 
\end{equation}
The coherent quasiparticle bands as measured in angle resolved photoemission spectroscopy (ARPES) is approximated by
\begin{equation} \label{eq:spectral-fnc}
	A(k,\omega) = \mathrm{Tr} \bm{A}(k,\omega) = \mathrm{Tr} (\bm{\mathcal{R}} \bm{\mathcal{R}}^{\dagger} \delta(\omega \bm{1} - \bm{\mathcal{H}}^{\mathrm{qp}}(k) ) ),
\end{equation}
so that the quasiparticle bands are given by the eigenenergies of $\hat{H}^{\mathrm{qp}}$, with their spectral weight renormalized by $\bm{\mathcal{R}} \bm{\mathcal{R}}^{\dagger} = \bm{\mathcal{R}}^{\dagger} \bm{\mathcal{R}}^{}$ ($\bm{\mathcal{R}}$ is Hermitian).

The spectral weight evaluated in the basis of the quasiparticle bands is 
\begin{align} 
	A(k,\omega) & = \sum_{p} [ \bm{U}^{\dagger}(k) \bm{\mathcal{R}}^{\dagger} \bm{\mathcal{R}}^{} \bm{U}^{}(k) ]_{pp} \delta(\omega - \xi_{p}^{\mathrm{qp}}(k)) \nonumber \\
	& = \sum_p Z_p^{\mathrm{qp}}(k) \delta(\omega - \xi_p^{\mathrm{qp}}(k))
\end{align}
where $\xi_{p}^{\mathrm{qp}}(k)$ are the eigenenergies of $\hat{H}^{\mathrm{qp}}$ (\cref{eq:ham-qp}) with $p$ indexing a band, $\bm{U}(k)$ is the single-particle unitary transformation that diagonalizes $\hat{H}^{\mathrm{qp}}$, and the quasiparticle weight of band $p$ is given by
\begin{equation} \label{eq:qp-weight-bands}
	Z_p^{\mathrm{qp}}(k) \equiv [\bm{U}^{\dagger}(k) \bm{\mathcal{R}}^{\dagger} \bm{\mathcal{R}}^{} \bm{U}^{}(k)]_{pp}.
\end{equation}

The quasiparticle weight matrix $\bm{Z}$ in the basis of the orbitals is related to the self-energy by
\begin{equation} \label{eq:qp-weight-orbitals}
	\bm{Z}_{\ell} \equiv \left( \bm{1} - \left. \frac{\partial \mathrm{Re} \bm{\Sigma}_{\ell}(\omega) }{\partial \omega} \right|_{\omega = 0} \right)^{-1} = \bm{\mathcal{R}}_{\ell}^{\dagger} \bm{\mathcal{R}}_{\ell}^{},
\end{equation}
and indciates the coherent spectral weight in the orbitals of cluster $\ell$ within the unit-cell by
\begin{equation} \label{eq:qp-weight-orbitals2}
	\int_{-\infty}^{\infty} \mathrm{d} \omega [\bm{A}_{\ell}]_{\alpha\beta}(k,\omega) = [\bm{\mathcal{R}}_{\ell}^{\dagger} \bm{\mathcal{R}}_{\ell}^{}]_{\alpha\beta} = [\bm{Z}_{\ell}]_{\alpha\beta}.
\end{equation}

\subsection{Implementation details for the decorated honeycomb lattice}

We implemented RISB within the \verb+TRIQS+ library \cite{SParcollet2015,SSeth2016} using two-site and three-site clusters (Fig.~2 of the main text). The $k$-integrals were evaluated using the linear tetrahedron method \cite{SBlochl1994}, and the impurity was solved using exact diagonalization using the Arnoldi method in \verb+ARPACK-NG+, interfaced by \verb+ezARPACK+ \cite{STRIQS-ARPACK}. We imposed the symmetry of $\hat{H}_i^{\mathrm{loc}}$ for the two-site and three-site clusters on the embedding state $|\Phi_i\rangle$ and the mean-field matrices $\bm{\lambda}$, $\bm{\mathcal{R}}$, $\bm{\lambda}^c$, and $\bm{\Delta}^p$. We assumed that the clusters were uniform.

The cluster choices within our study are the minimal size to observe the Mott insulators in their respective $t_g/t_k$ regimes because the insulators are extended phases from the molecular limit (\cref{sec:supp-dimer,sec:supp-trimer}). For example, a two-site cluster in the honeycomb-like regime ($t_g/t_k < 3/2$) is not sufficient to capture the intra-triangle correlations at two-thirds electron filling ($n=4/3$), and no spin-$1$ Mott insulator is found.

\section{Quasiparticle weight on the decorated honeycomb lattice}

\begin{figure}
	\centering
	\includegraphics[width=\columnwidth]{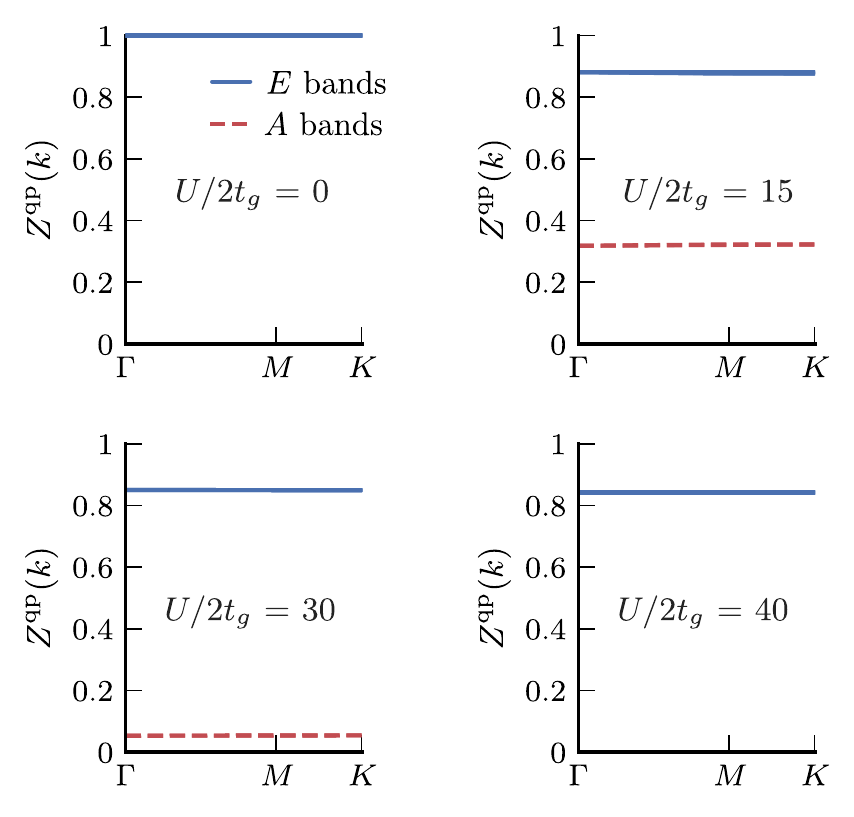} % final
	\caption{\label{fig:Z-qp} Quasiparticle weight $Z^{\mathrm{qp}}$ of the bands for $n=1/3$, $t_g / t_k = 0.4$. The set of $E$ ($A$) bands follows the labeling in Fig.~1b of the main text.
	}
\end{figure}

In \cref{fig:Z-qp} we show the quasiparticle weight $Z_p^{\mathrm{qp}}(k)$ (\cref{eq:qp-weight-bands}) of the bands for the trimer Mott insulator at $n=1/3$ using three-site clusters and assuming that the clusters are uniform on the lattice. At the $\Gamma$ point (Fig.~1b of the main text) $Z_p^{\mathrm{qp}}(k=\Gamma)$ for the six bands is equal to the quasiparticle weight in the basis of the molecular orbitals $\bm{Z} = \mathrm{diag}(Z_A,Z_A, Z_E,Z_E,Z_E,Z_E)$ (\cref{eq:qp-weight-orbitals,eq:qp-weight-orbitals2}). That is, for the lower bands $Z_A^{\mathrm{qp}}(k=\Gamma) = Z_A$ and for the upper bands $Z_E^{\mathrm{qp}}(k=\Gamma) = Z_E$. To a good approximation $Z_p^{\mathrm{qp}}(k) \approx Z_p^{\mathrm{qp}}(k=\Gamma) \equiv Z_{A,E}$. Because the quasiparticle weight $Z_A$ vanishes in the insulator, the spectrum of $\hat{H}^{\mathrm{qp}}$ (\cref{eq:ham-qp}) decouples the $A$ and $E$ bands. $Z_p^{\mathrm{qp}}(k)$ at $n=1/2,3/2$ (dimer Mott insulator) and $n=4/3$ (spin-$1$ Mott insulator) are analogous to the $n=1/3$ (trimer Mott insulator) case.

\begin{figure}
	\centering
	\includegraphics[width=\columnwidth]{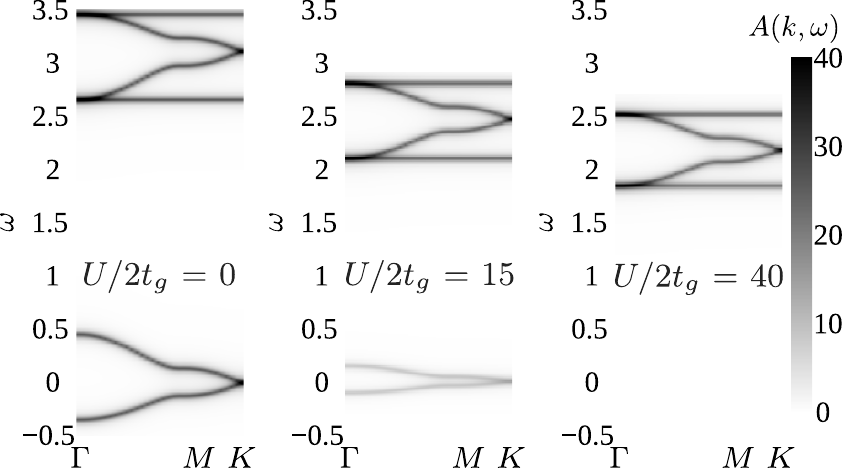} % final
	\caption{\label{fig:spectral-fn} Spectral function $A(k,\omega) = -\pi^{-1} \mathrm{Im} \mathrm{Tr} \bm{G}(k,\omega)$ for $n=1/3$, $t_g / t_k =0.4$. A broadening factor of $+i\eta = 0.025 i $ was used because the self-energy $\bm{\Sigma}(\omega)$ is purely real.
	}
\end{figure}

In \cref{fig:spectral-fn} we show the corresponding spectral function $A(k,\omega)$ (\cref{eq:spectral-fnc}) for the trimer Mott insulator at $n=1/3$. Interactions shift the quasiparticle bands, and the bands near the Fermi energy narrow. In the insulator the quasiparticle bands near the Fermi energy vanish. The behavior of $A(k,\omega)$ for the dimer Mott and spin-$1$ Mott insulators are analogous to the trimer Mott insulator.

\section{Hubbard model in the dimer basis} \label{sec:supp-dimer}

The Hamiltonian of the DHL clustered as dimers is given by
\begin{equation}
	\hat{H} \equiv \hat{H}^{\mathrm{kin}} + \sum_i \hat{H}_i^{\mathrm{loc}}
\end{equation}
with
\begin{align}
	\hat{H}^{\mathrm{kin}} & \equiv  - t_k \sum_{\langle i\alpha,j\alpha \rangle, \sigma} \hat{c}_{i\alpha\sigma}^{\dagger} \hat{c}_{j\alpha\sigma}^{}, \\
	\hat{H}_i^{\mathrm{loc}} & \equiv - t_g \sum_{\alpha\neq\beta ,\sigma} \hat{c}_{i\alpha\sigma}^{\dagger} \hat{c}_{i\beta\sigma}^{} + U\sum_{\alpha} \hat{n}_{i\alpha\uparrow} \hat{n}_{i\alpha\downarrow},
\end{align}
where $i,j$ label the dimer clusters, $\alpha \in \{1,2\}$ labels a site on a dimer, and $\sigma \in \{\uparrow,\downarrow\}$ is the electrons spin. We transform to the bonding (b) and  anti-bonding (a) basis of a dimer with
\begin{align}
	\hat{c}_{i 1 \sigma}^{\dagger} & \equiv \frac{1}{\sqrt{2}} \left( \hat{b}_{i b \sigma}^{\dagger} + \hat{b}_{i a \sigma}^{\dagger} \right), \nonumber \\
	\hat{c}_{i 2 \sigma}^{\dagger} & \equiv \frac{1}{\sqrt{2}} \left( \hat{b}_{i b \sigma}^{\dagger} - \hat{b}_{i a \sigma}^{\dagger} \right).
\end{align}

The local part of the Hamiltonian becomes
\begin{align}
	\hat{H}_i^{\mathrm{loc}} & = -t_g \sum_{\sigma} \left( \hat{b}_{i b \sigma}^{\dagger} \hat{b}_{i b \sigma}^{} - \hat{b}_{i a \sigma}^{\dagger} \hat{b}_{i a \sigma}^{}  \right) \nonumber \\
	& - \tilde{U} \vec{S}_i^2 + \frac{\tilde{U}}{4} \hat{N}_i^2 + \frac{\tilde{U}}{2} \hat{N}_i - \tilde{U} \sum_m \hat{n}_{im\uparrow} \hat{n}_{im\downarrow} \nonumber \\ 
	& + \tilde{U} \sum_{ \mathclap{m \neq n} } \hat{b}_{im\downarrow}^{\dagger} \hat{b}_{im\uparrow}^{\dagger} \hat{b}_{in\uparrow}^{} \hat{b}_{in \downarrow}^{},
\end{align}
where $\tilde{U} = U / 2$, $\hat{N}_i \equiv \sum_{m,\sigma} \hat{n}_{i m \sigma}$, $\vec{S}_i^2 = \sum_m \vec{S}_i \cdot \vec{S}_i$, and $\vec{S}_{im} = (S_{im}^x, S_{im}^y, S_{im}^z)$ is given by
\begin{equation}
	\vec{S}_{im} \equiv \frac{1}{2} \sum_{\sigma \sigma'} \hat{b}_{im \sigma}^{\dagger} \vec{\bm{\tau}}_{\sigma \sigma'} \hat{b}_{im \sigma'}^{},
\end{equation}
where $\vec{\bm{\tau}}$ is a vector of Pauli matrices.
The kinetic part of the Hamiltonian can be written as
\begin{equation}
	\hat{H}^{\mathrm{kin}} = \sum_{m} \hat{H}_m^{\mathrm{kagome}} + \sum_{m \neq n} \hat{H}_{mn}^{\mathrm{dimer}}.
\end{equation}
The first term is the tight-binding model on the kagome lattice given by
\begin{equation}
	\hat{H}_m^{\mathrm{kagome}} = \frac{1}{2} \sum_{k\sigma} \Psi_{km\sigma}^{\dagger} \bm{\mathcal{H}}_k^{\mathrm{kagome}} \vec{\Psi}_{km\sigma}^{},
\end{equation}
with $\vec{\Psi}_{km\sigma} \equiv (\hat{b}_{1km\sigma}, \hat{b}_{2km\sigma}, \hat{b}_{3km\sigma})^{\mathrm{T}}$ and
\begin{align}
	\bm{\mathcal{H}}_k^{\mathrm{kagome}} = -2 t_k \begin{pmatrix}
		0 & \cos k_1 & \cos k_2 \\
		\cos k_1 & 0 & \cos k_3 \\
		\cos k_2 & \cos k_3 & 0
	\end{pmatrix},
\end{align}
where $\ell = 1,2,3$ in $\hat{b}_{\ell k m\sigma}$ labels the three basis sites in the unit cell of the kagome lattice, and $k_j = \vec{k} \cdot \vec{a}_j$ with $\vec{a}_j$ denoting the three vectors to the nearest-neighbor sites in real-space.

The second term is a dimerization-like term in real-space between the (anti-)bonding orbitals on the two copies of the kagome lattice, given in $k$-space by
\begin{equation}
	\hat{H}_{mn}^{\mathrm{dimer}} = \frac{1}{2} \sum_{k\sigma} \vec{\Psi}_{km\sigma}^{\dagger} \bm{\mathcal{H}}_k^{\mathrm{dimer}} \vec{\Psi}_{kn\sigma}^{},
\end{equation}
with
\begin{align}
	\bm{\mathcal{H}}_k^{\mathrm{dimer}} = -2 t_k \begin{pmatrix}
		0 & i\sin k_1 & i\sin k_2 \\
		-i\sin k_1 & 0 & i\sin k_3 \\
		-i\sin k_2 & -i\sin k_3 & 0
	\end{pmatrix}.
\end{align}

In the limit where $t_g/t_k \rightarrow \infty$ the (anti-)bonding orbitals are infinitely separated in energy and the orbitals decouple. Electrons in either orbital are described by the single-orbital Hubbard model on the kagome lattice, given by
\begin{equation}
	\hat{H} = \hat{H}^{\mathrm{kagome}} + \tilde{U} \sum_{i} \hat{n}_{i \uparrow} \hat{n}_{i \downarrow}.
\end{equation}

\section{Hubbard model in the trimer basis} \label{sec:supp-trimer}

The Hamiltonian of the DHL clustered as dimers is given by
\begin{equation}
	\hat{H} \equiv \hat{H}^{\mathrm{kin}} + \sum_i \hat{H}_i^{\mathrm{loc}}
\end{equation}
with
\begin{align}
	\hat{H}^{\mathrm{kin}} & \equiv  - t_g \sum_{\langle i\alpha, j\alpha \rangle, \sigma} \hat{c}_{i\alpha\sigma}^{\dagger} \hat{c}_{j\alpha\sigma}^{} \\
	\hat{H}_i^{\mathrm{loc}} & \equiv - t_k \sum_{\alpha\neq\beta ,\sigma} \hat{c}_{i\alpha\sigma}^{\dagger} \hat{c}_{i\beta\sigma}^{} + U\sum_{\alpha} \hat{n}_{i\alpha\uparrow} \hat{n}_{i\alpha\downarrow},
\end{align}
where $i,j$ label the triangle clusters, $\alpha \in \{1,2,3\}$ labels a site on a triangle, and $\sigma \in \{\uparrow,\downarrow\}$ is the electrons spin. We transform to the molecular orbitals of a trimer with
\begin{equation}
	\hat{c}_{i \alpha \sigma}^{\dagger} \equiv \frac{1}{\sqrt{3}} \sum_m \hat{b}_{im\sigma}^{\dagger} e^{-i \phi (\alpha - 1) m},
\end{equation}
where $\phi = 2\pi/3$, $m \in \{-1,0,1\} \equiv \{E_1, A, E_2\}$ labels the trimer orbitals, and $m+m'$ is calculated modulo 3 shifted to the interval $[-1,1]$. 

The local part of the Hamiltonian becomes
\begin{align}
	& \hat{H}_i^{\mathrm{loc}} = -2t_k \sum_{m,\sigma} \cos(\phi m) \hat{b}_{im\sigma}^{\dagger} \hat{b}_{im\sigma}^{} \nonumber \\
	& + \tilde{U} \sum_m \hat{n}_{im\uparrow} \hat{n}_{im\downarrow} + \tilde{U} \sum_{m\neq n} \hat{n}_{im\uparrow} \hat{n}_{in\downarrow} \nonumber \\
	& - \tilde{U} \sum_{m\neq n} \vec{S}_{im} \cdot \vec{S}_{in}  +\tilde{U} \sum_{m\neq n} \hat{S}_{im}^z \hat{S}_{in}^z \nonumber \\
	& + \tilde{U} \sum_{m} \sum_{n \neq m} \sum_{p \neq m \neq n} \left( \hat{b}_{im \downarrow}^{\dagger} \hat{b}_{im \uparrow}^{\dagger} \hat{b}_{in \uparrow}^{} \hat{b}_{ip \downarrow}^{} + \mathrm{H.c.} \right) \nonumber \\
	& = -2t_k \sum_{m,\sigma} \cos(\phi m) \hat{b}_{im\sigma}^{\dagger} \hat{b}_{im\sigma}^{} \nonumber \\
	& - \tilde{U} \vec{S}_i^2 + \frac{\tilde{U}}{4}  \hat{N}_i^2 + \frac{\tilde{U}}{2}  \hat{N}_i - \tilde{U} \sum_m \hat{n}_{im\uparrow} \hat{n}_{im\downarrow} \nonumber \\
	& + \tilde{U} \sum_{m} \sum_{n \neq m} \sum_{p \neq m \neq n} \left( \hat{b}_{im \downarrow}^{\dagger} \hat{b}_{im \uparrow}^{\dagger} \hat{b}_{in \uparrow}^{} \hat{b}_{ip \downarrow}^{} + \mathrm{H.c.} \right),
\end{align}
where $\tilde{U} = U/3$.
The kinetic part of the Hamiltonian can be written as
\begin{equation}
	\hat{H}^{\mathrm{kin}} = \sum_{m} \hat{H}_{m}^{\mathrm{honeycomb}} + \sum_{m \neq n} \hat{H}_{mn}^{\mathrm{trimer}}.
\end{equation}
The first term is the tight-binding model on the honeycomb lattice given by
\begin{equation}
	\hat{H}_m^{\mathrm{honeycomb}} = \frac{1}{3} \sum_{k\sigma} \vec{\Psi}_{km\sigma}^{\dagger} \bm{\mathcal{H}}_{k}^{\mathrm{honeycomb}} \vec{\Psi}_{km\sigma}^{},
\end{equation}
with $\vec{\Psi}_{km\sigma} \equiv (\hat{b}_{1km\sigma}, \hat{b}_{2km\sigma})^{\mathrm{T}}$ and
\begin{align}
	\bm{\mathcal{H}}_k^{\mathrm{honeycomb}} = - t_g \begin{pmatrix}
		0 & \Delta_k \\
		\Delta_k^* & 0
	\end{pmatrix},
\end{align}
where $\ell = 1,2$ in $\hat{b}_{\ell km\sigma}$ labels the two basis sites in the unit cell of the honeycomb lattice, and $\Delta_k = \sum_j e^{i \vec{k} \cdot \vec{a}_j}$ with $\vec{a}_j$ denoting the three vectors to the nearest-neighbor sites in real-space.

The second term between trimer orbitals takes a similar form but has an additional phase dependence, given in $k$-space by
\begin{equation}
	\hat{H}_{mn}^{\mathrm{trimer}} = \frac{1}{3} \sum_{k\sigma} \vec{\Psi}_{km\sigma}^{\dagger} \bm{\mathcal{H}}_{k}^{\mathrm{trimer}} \vec{\Psi}_{kn\sigma}^{},
\end{equation}
with
\begin{align}
	\bm{\mathcal{H}}_k^{\mathrm{trimer}} = - t_g \begin{pmatrix}
		0 & \delta_k \\
		\delta_k^* & 0
	\end{pmatrix},
\end{align}
where $\delta_k = \sum_{j=1}^{3} e^{i (\vec{k} \cdot \vec{a}_j + (j-1) \phi)}$.

In the limit where $t_g / t_k \rightarrow 0$ the $A$ and $E$ orbitals are infinitely separated in energy and we can model the electrons with two decoupled Hamiltonians. Electrons in the $A$ orbitals are described by the Hubbard model on the honeycomb lattice
\begin{equation}
	\hat{H}^A = \hat{H}^{\mathrm{honeycomb}} + \tilde{U} \sum_i \hat{n}_{i\uparrow} \hat{n}_{i\downarrow}.
\end{equation}
Electrons in the degenerate $E$ orbitals are described by a two-orbital Hubbard model on a honeycomb-like lattice given by
\begin{align}
	\hat{H}^E & = \sum_m \hat{H}_m^{\mathrm{honeycomb}} + \sum_{m\neq n} \hat{H}_{mn}^{\mathrm{trimer}} \nonumber \\
	& - \tilde{U} \vec{S}_i^2 + \frac{\tilde{U}}{4}  \hat{N}_i^2 + \frac{\tilde{U}}{2}  \hat{N}_i - \tilde{U} \sum_m \hat{n}_{im\uparrow} \hat{n}_{im\downarrow}, 
\end{align}
where $m \in \{E_1, E_2\}$.

\section{Magnetic solutions}

\begin{figure}
	\centering
	\includegraphics[width=\columnwidth]{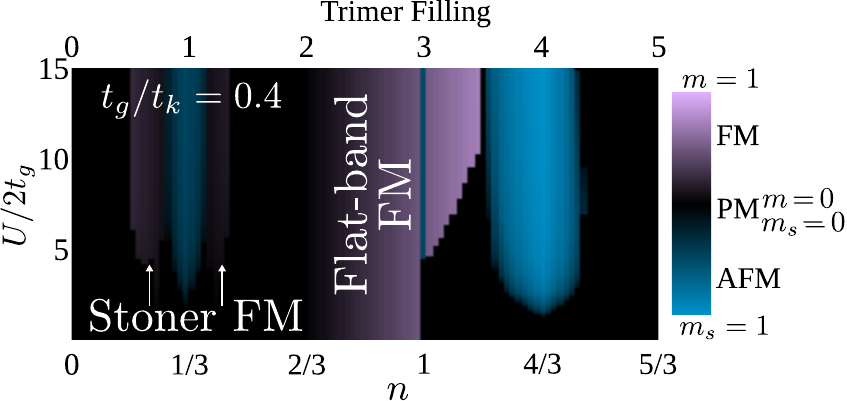} % final
	\caption{\label{fig:heatmaps-trimer-mag}
		Magnetic ordering between triangular clusters of the DHL in the honeycomb-like regime for the triangle clusters (Fig.2~d-f of the main text).
		AFM is identified with $m = 0$, $m_s \neq 0$, FM with $m \neq 0$, $m_s = 0$, and paramagnetism (PM) with $m = m_s = 0$, where
		$m \equiv \sum_i \sum_{\alpha=1}^3 \langle \hat{S}_{i\alpha}^z \rangle / 2 \mathcal{N}$, $m_s \equiv \sum_i (-1)^i \sum_{\alpha=1}^3 \langle \hat{S}_{i\alpha}^z \rangle / 2 \mathcal{N}$, $\hat{S}_{i\alpha}^z = \frac{1}{2} ( \hat{c}_{i\alpha\uparrow}^{\dagger} \hat{c}_{i\alpha\uparrow}^{} - \hat{c}_{i\alpha\downarrow}^{\dagger} \hat{c}_{i\alpha\downarrow}^{})$, and $\mathcal{N}$ is the number of unit cells.
	}
\end{figure}

The magnetization $m$ and staggered magnetization $m_s$ are plotted in \cref{fig:heatmaps-trimer-mag}. 
Antiferromagnetism (AFM) dominates in the vicinity of the insulators (trimer Mott at $n = 1/3$ and spin-1 Mott at $n = 4/3$), and exactly at half-filling ($n=1$). Ferromagnetism (FM) is favored near the diverging density of states at the van Hove singularities ($n=1/4, 5/12$), and the flat band ($2/3 < n < 1$).

\subsection{Broken $\mathcal{C}_3$ symmetry at half-filling}

\begin{figure}
	\centering
	\includegraphics[width=\columnwidth]{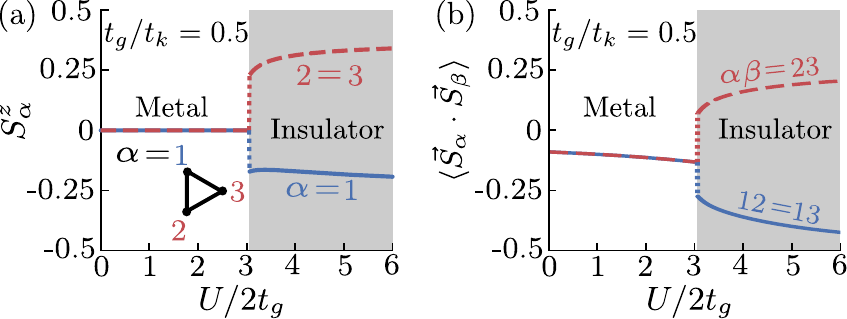} % final
	\caption{\label{fig:dh-half-filling-broken-c3}
		The $\mathcal{C}_3$ rotational symmetry of a triangle is broken for $t_g / t_k < 0.9-1$ at half-filling.
		(a) Spin magnetization $S^z$ of each site and (b) spin-exchange between adjacent sites within a triangle.
	}
\end{figure}

The magnetization per site $S_{\alpha}^z$ and intra-triangle spin-exchange $\vec{S}_{\alpha} \cdot \vec{S}_{\beta}$, for $\alpha \neq \beta$, are plotted in \cref{fig:dh-half-filling-broken-c3} at half-filling ($n=1$). For $t_g / t_k > 0.9-1$ the ground state is paramagnetic and favors spin-singlets along the $t_g$ bonds (Fig.~4b of the main text). For $t_g / t_k < 0.9-1$ the paramagnetic solutions favors spin-singlets along the $t_k$ bonds (Fig.~3b of the main text), and there is an instability to an antiferromagnetic insulator (\cref{fig:heatmaps-trimer-mag}) with electrons localized to a site. The frustration of the lattice makes it difficult to evenly arrange the three spins on a triangle, and as a consequence the ground state breaks the $\mathcal{C}_3$ rotational symmetry of the lattice.

%\bibliography{bibliography}

\begin{thebibliography}{85}%
	\makeatletter
	\providecommand \@ifxundefined [1]{%
		\@ifx{#1\undefined}
	}%
	\providecommand \@ifnum [1]{%
		\ifnum #1\expandafter \@firstoftwo
		\else \expandafter \@secondoftwo
		\fi
	}%
	\providecommand \@ifx [1]{%
		\ifx #1\expandafter \@firstoftwo
		\else \expandafter \@secondoftwo
		\fi
	}%
	\providecommand \natexlab [1]{#1}%
	\providecommand \enquote  [1]{``#1''}%
	\providecommand \bibnamefont  [1]{#1}%
	\providecommand \bibfnamefont [1]{#1}%
	\providecommand \citenamefont [1]{#1}%
	\providecommand \href@noop [0]{\@secondoftwo}%
	\providecommand \href [0]{\begingroup \@sanitize@url \@href}%
	\providecommand \@href[1]{\@@startlink{#1}\@@href}%
	\providecommand \@@href[1]{\endgroup#1\@@endlink}%
	\providecommand \@sanitize@url [0]{\catcode `\\12\catcode `\$12\catcode
		`\&12\catcode `\#12\catcode `\^12\catcode `\_12\catcode `\%12\relax}%
	\providecommand \@@startlink[1]{}%
	\providecommand \@@endlink[0]{}%
	\providecommand \url  [0]{\begingroup\@sanitize@url \@url }%
	\providecommand \@url [1]{\endgroup\@href {#1}{\urlprefix }}%
	\providecommand \urlprefix  [0]{URL }%
	\providecommand \Eprint [0]{\href }%
	\providecommand \doibase [0]{https://doi.org/}%
	\providecommand \selectlanguage [0]{\@gobble}%
	\providecommand \bibinfo  [0]{\@secondoftwo}%
	\providecommand \bibfield  [0]{\@secondoftwo}%
	\providecommand \translation [1]{[#1]}%
	\providecommand \BibitemOpen [0]{}%
	\providecommand \bibitemStop [0]{}%
	\providecommand \bibitemNoStop [0]{.\EOS\space}%
	\providecommand \EOS [0]{\spacefactor3000\relax}%
	\providecommand \BibitemShut  [1]{\csname bibitem#1\endcsname}%
	\let\auto@bib@innerbib\@empty
	%</preamble>
	\bibitem [{\citenamefont {Zheng}\ \emph {et~al.}(2007)\citenamefont {Zheng},
		\citenamefont {Tong}, \citenamefont {Xue}, \citenamefont {Zhang},
		\citenamefont {Chen}, \citenamefont {Grandjean},\ and\ \citenamefont
		{Long}}]{Zheng2007}%
	\BibitemOpen
	\bibfield  {author} {\bibinfo {author} {\bibfnamefont {Y.-Z.}\ \bibnamefont
			{Zheng}}, \bibinfo {author} {\bibfnamefont {M.-L.}\ \bibnamefont {Tong}},
		\bibinfo {author} {\bibfnamefont {W.}~\bibnamefont {Xue}}, \bibinfo {author}
		{\bibfnamefont {W.-X.}\ \bibnamefont {Zhang}}, \bibinfo {author}
		{\bibfnamefont {X.-M.}\ \bibnamefont {Chen}}, \bibinfo {author}
		{\bibfnamefont {F.}~\bibnamefont {Grandjean}},\ and\ \bibinfo {author}
		{\bibfnamefont {G.}~\bibnamefont {Long}},\ }\href
	{https://doi.org/10.1002/anie.200701954} {\bibfield  {journal} {\bibinfo
			{journal} {Angew. Chem., Int. Ed.}\ }\textbf {\bibinfo {volume} {46}},\
		\bibinfo {pages} {6076} (\bibinfo {year} {2007})}\BibitemShut {NoStop}%
	\bibitem [{\citenamefont {Bao}\ \emph {et~al.}(2015)\citenamefont {Bao},
		\citenamefont {Liu}, \citenamefont {Ma}, \citenamefont {Meng}, \citenamefont
		{Tang}, \citenamefont {Sun}, \citenamefont {Zhai}, \citenamefont {Jiang},
		\citenamefont {Bai}, \citenamefont {Feng}, \citenamefont {Xu},\ and\
		\citenamefont {Cao}}]{Bao2015}%
	\BibitemOpen
	\bibfield  {author} {\bibinfo {author} {\bibfnamefont {J.-K.}\ \bibnamefont
			{Bao}}, \bibinfo {author} {\bibfnamefont {J.-Y.}\ \bibnamefont {Liu}},
		\bibinfo {author} {\bibfnamefont {C.-W.}\ \bibnamefont {Ma}}, \bibinfo
		{author} {\bibfnamefont {Z.-H.}\ \bibnamefont {Meng}}, \bibinfo {author}
		{\bibfnamefont {Z.-T.}\ \bibnamefont {Tang}}, \bibinfo {author}
		{\bibfnamefont {Y.-L.}\ \bibnamefont {Sun}}, \bibinfo {author} {\bibfnamefont
			{H.-F.}\ \bibnamefont {Zhai}}, \bibinfo {author} {\bibfnamefont
			{H.}~\bibnamefont {Jiang}}, \bibinfo {author} {\bibfnamefont
			{H.}~\bibnamefont {Bai}}, \bibinfo {author} {\bibfnamefont {C.-M.}\
			\bibnamefont {Feng}}, \bibinfo {author} {\bibfnamefont {Z.-A.}\ \bibnamefont
			{Xu}},\ and\ \bibinfo {author} {\bibfnamefont {G.-H.}\ \bibnamefont {Cao}},\
	}\href {https://doi.org/10.1103/PhysRevX.5.011013} {\bibfield  {journal}
		{\bibinfo  {journal} {Phys. Rev. X}\ }\textbf {\bibinfo {volume} {5}},\
		\bibinfo {pages} {011013} (\bibinfo {year} {2015})}\BibitemShut {NoStop}%
	\bibitem [{\citenamefont {Nguyen}\ \emph {et~al.}(2018)\citenamefont {Nguyen},
		\citenamefont {Halloran}, \citenamefont {Xie}, \citenamefont {Kong},
		\citenamefont {Broholm},\ and\ \citenamefont {Cava}}]{Nguyen2018}%
	\BibitemOpen
	\bibfield  {author} {\bibinfo {author} {\bibfnamefont {L.~T.}\ \bibnamefont
			{Nguyen}}, \bibinfo {author} {\bibfnamefont {T.}~\bibnamefont {Halloran}},
		\bibinfo {author} {\bibfnamefont {W.}~\bibnamefont {Xie}}, \bibinfo {author}
		{\bibfnamefont {T.}~\bibnamefont {Kong}}, \bibinfo {author} {\bibfnamefont
			{C.~L.}\ \bibnamefont {Broholm}},\ and\ \bibinfo {author} {\bibfnamefont
			{R.~J.}\ \bibnamefont {Cava}},\ }\href
	{https://link.aps.org/doi/10.1103/PhysRevMaterials.2.054414} {\bibfield
		{journal} {\bibinfo  {journal} {Phys. Rev. Mater.}\ }\textbf {\bibinfo
			{volume} {2}},\ \bibinfo {pages} {054414} (\bibinfo {year}
		{2018})}\BibitemShut {NoStop}%
	\bibitem [{\citenamefont {Taniguchi}\ \emph {et~al.}(1995)\citenamefont
		{Taniguchi}, \citenamefont {Nishikawa}, \citenamefont {Yasui}, \citenamefont
		{Kobayashi}, \citenamefont {Sato}, \citenamefont {Nishioka}, \citenamefont
		{Kontani},\ and\ \citenamefont {Sano}}]{Taniguchi1995}%
	\BibitemOpen
	\bibfield  {author} {\bibinfo {author} {\bibfnamefont {S.}~\bibnamefont
			{Taniguchi}}, \bibinfo {author} {\bibfnamefont {T.}~\bibnamefont
			{Nishikawa}}, \bibinfo {author} {\bibfnamefont {Y.}~\bibnamefont {Yasui}},
		\bibinfo {author} {\bibfnamefont {Y.}~\bibnamefont {Kobayashi}}, \bibinfo
		{author} {\bibfnamefont {M.}~\bibnamefont {Sato}}, \bibinfo {author}
		{\bibfnamefont {T.}~\bibnamefont {Nishioka}}, \bibinfo {author}
		{\bibfnamefont {M.}~\bibnamefont {Kontani}},\ and\ \bibinfo {author}
		{\bibfnamefont {K.}~\bibnamefont {Sano}},\ }\href
	{https://doi.org/10.1143/JPSJ.64.2758} {\bibfield  {journal} {\bibinfo
			{journal} {J. Phys. Soc. Jpn.}\ }\textbf {\bibinfo {volume} {64}},\ \bibinfo
		{pages} {2758} (\bibinfo {year} {1995})}\BibitemShut {NoStop}%
	\bibitem [{\citenamefont {Ye}\ \emph {et~al.}(2011)\citenamefont {Ye},
		\citenamefont {Chi}, \citenamefont {Bao}, \citenamefont {Wang}, \citenamefont
		{Ying}, \citenamefont {Chen}, \citenamefont {Wang}, \citenamefont {Dong},\
		and\ \citenamefont {Fang}}]{Ye2011}%
	\BibitemOpen
	\bibfield  {author} {\bibinfo {author} {\bibfnamefont {F.}~\bibnamefont
			{Ye}}, \bibinfo {author} {\bibfnamefont {S.}~\bibnamefont {Chi}}, \bibinfo
		{author} {\bibfnamefont {W.}~\bibnamefont {Bao}}, \bibinfo {author}
		{\bibfnamefont {X.~F.}\ \bibnamefont {Wang}}, \bibinfo {author}
		{\bibfnamefont {J.~J.}\ \bibnamefont {Ying}}, \bibinfo {author}
		{\bibfnamefont {X.~H.}\ \bibnamefont {Chen}}, \bibinfo {author}
		{\bibfnamefont {H.~D.}\ \bibnamefont {Wang}}, \bibinfo {author}
		{\bibfnamefont {C.~H.}\ \bibnamefont {Dong}},\ and\ \bibinfo {author}
		{\bibfnamefont {M.}~\bibnamefont {Fang}},\ }\href
	{https://doi.org/10.1103/PhysRevLett.107.137003} {\bibfield  {journal}
		{\bibinfo  {journal} {Phys. Rev. Lett.}\ }\textbf {\bibinfo {volume} {107}},\
		\bibinfo {pages} {137003} (\bibinfo {year} {2011})}\BibitemShut {NoStop}%
	\bibitem [{\citenamefont {Bao}\ \emph {et~al.}(2011)\citenamefont {Bao},
		\citenamefont {Huang}, \citenamefont {Chen}, \citenamefont {Wang},
		\citenamefont {He},\ and\ \citenamefont {Qiu}}]{Bao2011}%
	\BibitemOpen
	\bibfield  {author} {\bibinfo {author} {\bibfnamefont {W.}~\bibnamefont
			{Bao}}, \bibinfo {author} {\bibfnamefont {Q.-Z.}\ \bibnamefont {Huang}},
		\bibinfo {author} {\bibfnamefont {G.-F.}\ \bibnamefont {Chen}}, \bibinfo
		{author} {\bibfnamefont {D.-M.}\ \bibnamefont {Wang}}, \bibinfo {author}
		{\bibfnamefont {J.-B.}\ \bibnamefont {He}},\ and\ \bibinfo {author}
		{\bibfnamefont {Y.-M.}\ \bibnamefont {Qiu}},\ }\href
	{https://doi.org/10.1088/0256-307x/28/8/086104} {\bibfield  {journal}
		{\bibinfo  {journal} {Chin. Phys. Lett.}\ }\textbf {\bibinfo {volume} {28}},\
		\bibinfo {pages} {086104} (\bibinfo {year} {2011})}\BibitemShut {NoStop}%
	\bibitem [{\citenamefont {Jacko}\ \emph {et~al.}(2015)\citenamefont {Jacko},
		\citenamefont {Janani}, \citenamefont {Koepernik},\ and\ \citenamefont
		{Powell}}]{Jacko2015}%
	\BibitemOpen
	\bibfield  {author} {\bibinfo {author} {\bibfnamefont {A.~C.}\ \bibnamefont
			{Jacko}}, \bibinfo {author} {\bibfnamefont {C.}~\bibnamefont {Janani}},
		\bibinfo {author} {\bibfnamefont {K.}~\bibnamefont {Koepernik}},\ and\
		\bibinfo {author} {\bibfnamefont {B.~J.}\ \bibnamefont {Powell}},\ }\href
	{http://link.aps.org/doi/10.1103/PhysRevB.91.125140} {\bibfield  {journal}
		{\bibinfo  {journal} {Phys. Rev. B}\ }\textbf {\bibinfo {volume} {91}},\
		\bibinfo {pages} {125140} (\bibinfo {year} {2015})}\BibitemShut {NoStop}%
	\bibitem [{\citenamefont {Shuku}\ \emph {et~al.}(2018)\citenamefont {Shuku},
		\citenamefont {Mizuno}, \citenamefont {Ushiroguchi}, \citenamefont {Hyun},
		\citenamefont {Ryu}, \citenamefont {An}, \citenamefont {Kwon}, \citenamefont
		{Park}, \citenamefont {Tsuchiizu},\ and\ \citenamefont {Awaga}}]{Shuku2018}%
	\BibitemOpen
	\bibfield  {author} {\bibinfo {author} {\bibfnamefont {Y.}~\bibnamefont
			{Shuku}}, \bibinfo {author} {\bibfnamefont {A.}~\bibnamefont {Mizuno}},
		\bibinfo {author} {\bibfnamefont {R.}~\bibnamefont {Ushiroguchi}}, \bibinfo
		{author} {\bibfnamefont {C.~S.}\ \bibnamefont {Hyun}}, \bibinfo {author}
		{\bibfnamefont {Y.~J.}\ \bibnamefont {Ryu}}, \bibinfo {author} {\bibfnamefont
			{B.-K.}\ \bibnamefont {An}}, \bibinfo {author} {\bibfnamefont {J.~E.}\
			\bibnamefont {Kwon}}, \bibinfo {author} {\bibfnamefont {S.~Y.}\ \bibnamefont
			{Park}}, \bibinfo {author} {\bibfnamefont {M.}~\bibnamefont {Tsuchiizu}},\
		and\ \bibinfo {author} {\bibfnamefont {K.}~\bibnamefont {Awaga}},\ }\href
	{https://doi.org/10.1039/C8CC00753E} {\bibfield  {journal} {\bibinfo
			{journal} {Chem. Commun.}\ }\textbf {\bibinfo {volume} {54}},\ \bibinfo
		{pages} {3815} (\bibinfo {year} {2018})}\BibitemShut {NoStop}%
	\bibitem [{\citenamefont {Murase}\ \emph
		{et~al.}(2017{\natexlab{a}})\citenamefont {Murase}, \citenamefont {Leong},\
		and\ \citenamefont {D’Alessandro}}]{Murase2017}%
	\BibitemOpen
	\bibfield  {author} {\bibinfo {author} {\bibfnamefont {R.}~\bibnamefont
			{Murase}}, \bibinfo {author} {\bibfnamefont {C.~F.}\ \bibnamefont {Leong}},\
		and\ \bibinfo {author} {\bibfnamefont {D.~M.}\ \bibnamefont
			{D’Alessandro}},\ }\href {https://doi.org/10.1021/acs.inorgchem.7b02090}
	{\bibfield  {journal} {\bibinfo  {journal} {Inorg. Chem.}\ }\textbf {\bibinfo
			{volume} {56}},\ \bibinfo {pages} {14373} (\bibinfo {year}
		{2017}{\natexlab{a}})}\BibitemShut {NoStop}%
	\bibitem [{\citenamefont {Murase}\ \emph
		{et~al.}(2017{\natexlab{b}})\citenamefont {Murase}, \citenamefont {Abrahams},
		\citenamefont {D’Alessandro}, \citenamefont {Davies}, \citenamefont
		{Hudson}, \citenamefont {Jameson}, \citenamefont {Moubaraki}, \citenamefont
		{Murray}, \citenamefont {Robson},\ and\ \citenamefont
		{Sutton}}]{Murase2017b}%
	\BibitemOpen
	\bibfield  {author} {\bibinfo {author} {\bibfnamefont {R.}~\bibnamefont
			{Murase}}, \bibinfo {author} {\bibfnamefont {B.~F.}\ \bibnamefont
			{Abrahams}}, \bibinfo {author} {\bibfnamefont {D.~M.}\ \bibnamefont
			{D’Alessandro}}, \bibinfo {author} {\bibfnamefont {C.~G.}\ \bibnamefont
			{Davies}}, \bibinfo {author} {\bibfnamefont {T.~A.}\ \bibnamefont {Hudson}},
		\bibinfo {author} {\bibfnamefont {G.~N.~L.}\ \bibnamefont {Jameson}},
		\bibinfo {author} {\bibfnamefont {B.}~\bibnamefont {Moubaraki}}, \bibinfo
		{author} {\bibfnamefont {K.~S.}\ \bibnamefont {Murray}}, \bibinfo {author}
		{\bibfnamefont {R.}~\bibnamefont {Robson}},\ and\ \bibinfo {author}
		{\bibfnamefont {A.~L.}\ \bibnamefont {Sutton}},\ }\href
	{https://doi.org/10.1021/acs.inorgchem.7b01038} {\bibfield  {journal}
		{\bibinfo  {journal} {Inorg. Chem.}\ }\textbf {\bibinfo {volume} {56}},\
		\bibinfo {pages} {9025} (\bibinfo {year} {2017}{\natexlab{b}})}\BibitemShut
	{NoStop}%
	\bibitem [{\citenamefont {Kingsbury}\ \emph {et~al.}(2017)\citenamefont
		{Kingsbury}, \citenamefont {Abrahams}, \citenamefont {D’Alessandro},
		\citenamefont {Hudson}, \citenamefont {Murase}, \citenamefont {Robson},\ and\
		\citenamefont {White}}]{Kingsbury2017}%
	\BibitemOpen
	\bibfield  {author} {\bibinfo {author} {\bibfnamefont {C.~J.}\ \bibnamefont
			{Kingsbury}}, \bibinfo {author} {\bibfnamefont {B.~F.}\ \bibnamefont
			{Abrahams}}, \bibinfo {author} {\bibfnamefont {D.~M.}\ \bibnamefont
			{D’Alessandro}}, \bibinfo {author} {\bibfnamefont {T.~A.}\ \bibnamefont
			{Hudson}}, \bibinfo {author} {\bibfnamefont {R.}~\bibnamefont {Murase}},
		\bibinfo {author} {\bibfnamefont {R.}~\bibnamefont {Robson}},\ and\ \bibinfo
		{author} {\bibfnamefont {K.~F.}\ \bibnamefont {White}},\ }\href
	{https://doi.org/10.1021/acs.cgd.6b01886} {\bibfield  {journal} {\bibinfo
			{journal} {Cryst. Growth Des.}\ }\textbf {\bibinfo {volume} {17}},\ \bibinfo
		{pages} {1465} (\bibinfo {year} {2017})}\BibitemShut {NoStop}%
	\bibitem [{\citenamefont {Jeon}\ \emph {et~al.}(2015)\citenamefont {Jeon},
		\citenamefont {Negru}, \citenamefont {Van~Duyne},\ and\ \citenamefont
		{Harris}}]{Jeon2015}%
	\BibitemOpen
	\bibfield  {author} {\bibinfo {author} {\bibfnamefont {I.-R.}\ \bibnamefont
			{Jeon}}, \bibinfo {author} {\bibfnamefont {B.}~\bibnamefont {Negru}},
		\bibinfo {author} {\bibfnamefont {R.~P.}\ \bibnamefont {Van~Duyne}},\ and\
		\bibinfo {author} {\bibfnamefont {T.~D.}\ \bibnamefont {Harris}},\ }\href
	{https://doi.org/10.1021/jacs.5b10382} {\bibfield  {journal} {\bibinfo
			{journal} {J. Am. Chem. Soc.}\ }\textbf {\bibinfo {volume} {137}},\ \bibinfo
		{pages} {15699} (\bibinfo {year} {2015})}\BibitemShut {NoStop}%
	\bibitem [{\citenamefont {Darago}\ \emph {et~al.}(2015)\citenamefont {Darago},
		\citenamefont {Aubrey}, \citenamefont {Yu}, \citenamefont {Gonzalez},\ and\
		\citenamefont {Long}}]{Darago2015}%
	\BibitemOpen
	\bibfield  {author} {\bibinfo {author} {\bibfnamefont {L.~E.}\ \bibnamefont
			{Darago}}, \bibinfo {author} {\bibfnamefont {M.~L.}\ \bibnamefont {Aubrey}},
		\bibinfo {author} {\bibfnamefont {C.~J.}\ \bibnamefont {Yu}}, \bibinfo
		{author} {\bibfnamefont {M.~I.}\ \bibnamefont {Gonzalez}},\ and\ \bibinfo
		{author} {\bibfnamefont {J.~R.}\ \bibnamefont {Long}},\ }\href
	{https://doi.org/10.1021/jacs.5b10385} {\bibfield  {journal} {\bibinfo
			{journal} {J. Am. Chem. Soc.}\ }\textbf {\bibinfo {volume} {137}},\ \bibinfo
		{pages} {15703} (\bibinfo {year} {2015})}\BibitemShut {NoStop}%
	\bibitem [{\citenamefont {DeGayner}\ \emph {et~al.}(2017)\citenamefont
		{DeGayner}, \citenamefont {Jeon}, \citenamefont {Sun}, \citenamefont
		{Dincă},\ and\ \citenamefont {Harris}}]{DeGayner2017}%
	\BibitemOpen
	\bibfield  {author} {\bibinfo {author} {\bibfnamefont {J.~A.}\ \bibnamefont
			{DeGayner}}, \bibinfo {author} {\bibfnamefont {I.-R.}\ \bibnamefont {Jeon}},
		\bibinfo {author} {\bibfnamefont {L.}~\bibnamefont {Sun}}, \bibinfo {author}
		{\bibfnamefont {M.}~\bibnamefont {Dincă}},\ and\ \bibinfo {author}
		{\bibfnamefont {T.~D.}\ \bibnamefont {Harris}},\ }\href
	{https://doi.org/10.1021/jacs.7b00705} {\bibfield  {journal} {\bibinfo
			{journal} {J. Am. Chem. Soc.}\ }\textbf {\bibinfo {volume} {139}},\ \bibinfo
		{pages} {4175} (\bibinfo {year} {2017})}\BibitemShut {NoStop}%
	\bibitem [{\citenamefont {Henling}\ and\ \citenamefont
		{Marsh}(2014)}]{Henling2014}%
	\BibitemOpen
	\bibfield  {author} {\bibinfo {author} {\bibfnamefont {L.~M.}\ \bibnamefont
			{Henling}}\ and\ \bibinfo {author} {\bibfnamefont {R.~E.}\ \bibnamefont
			{Marsh}},\ }\href {https://doi.org/10.1107/S2053229614017549} {\bibfield
		{journal} {\bibinfo  {journal} {Acta Crystallogr., Sect. C: Struct. Chem.}\
		}\textbf {\bibinfo {volume} {70}},\ \bibinfo {pages} {834} (\bibinfo {year}
		{2014})},\ \bibinfo {note} {{C}SD-FAZGIY}\BibitemShut {NoStop}%
	\bibitem [{\citenamefont {Henline}\ \emph {et~al.}(2014)\citenamefont
		{Henline}, \citenamefont {Wang}, \citenamefont {Pike}, \citenamefont {Ahern},
		\citenamefont {Sousa}, \citenamefont {Patterson}, \citenamefont {Kerr},\ and\
		\citenamefont {Cahill}}]{Henline2014}%
	\BibitemOpen
	\bibfield  {author} {\bibinfo {author} {\bibfnamefont {K.~M.}\ \bibnamefont
			{Henline}}, \bibinfo {author} {\bibfnamefont {C.}~\bibnamefont {Wang}},
		\bibinfo {author} {\bibfnamefont {R.~D.}\ \bibnamefont {Pike}}, \bibinfo
		{author} {\bibfnamefont {J.~C.}\ \bibnamefont {Ahern}}, \bibinfo {author}
		{\bibfnamefont {B.}~\bibnamefont {Sousa}}, \bibinfo {author} {\bibfnamefont
			{H.~H.}\ \bibnamefont {Patterson}}, \bibinfo {author} {\bibfnamefont {A.~T.}\
			\bibnamefont {Kerr}},\ and\ \bibinfo {author} {\bibfnamefont {C.~L.}\
			\bibnamefont {Cahill}},\ }\href {https://doi.org/10.1021/cg500005p}
	{\bibfield  {journal} {\bibinfo  {journal} {Cryst. Growth Des.}\ }\textbf
		{\bibinfo {volume} {14}},\ \bibinfo {pages} {1449} (\bibinfo {year}
		{2014})}\BibitemShut {NoStop}%
	\bibitem [{\citenamefont {Polunin}\ \emph {et~al.}(2015)\citenamefont
		{Polunin}, \citenamefont {Dorofeeva}, \citenamefont {Baranchikov},
		\citenamefont {Ivanov}, \citenamefont {Gavrilenko}, \citenamefont {Kiskin},
		\citenamefont {Eremenko}, \citenamefont {Novotortsev},\ and\ \citenamefont
		{Kolotilov}}]{Polunin2015}%
	\BibitemOpen
	\bibfield  {author} {\bibinfo {author} {\bibfnamefont {R.~A.}\ \bibnamefont
			{Polunin}}, \bibinfo {author} {\bibfnamefont {V.~N.}\ \bibnamefont
			{Dorofeeva}}, \bibinfo {author} {\bibfnamefont {A.~E.}\ \bibnamefont
			{Baranchikov}}, \bibinfo {author} {\bibfnamefont {V.~K.}\ \bibnamefont
			{Ivanov}}, \bibinfo {author} {\bibfnamefont {K.~S.}\ \bibnamefont
			{Gavrilenko}}, \bibinfo {author} {\bibfnamefont {M.~A.}\ \bibnamefont
			{Kiskin}}, \bibinfo {author} {\bibfnamefont {I.~L.}\ \bibnamefont
			{Eremenko}}, \bibinfo {author} {\bibfnamefont {V.~M.}\ \bibnamefont
			{Novotortsev}},\ and\ \bibinfo {author} {\bibfnamefont {S.~V.}\ \bibnamefont
			{Kolotilov}},\ }\href {https://doi.org/10.1134/S1070328415060056} {\bibfield
		{journal} {\bibinfo  {journal} {Russ. J. Coord. Chem.}\ }\textbf {\bibinfo
			{volume} {41}},\ \bibinfo {pages} {353} (\bibinfo {year} {2015})}\BibitemShut
	{NoStop}%
	\bibitem [{\citenamefont {Kalmutzki}\ \emph {et~al.}(2018)\citenamefont
		{Kalmutzki}, \citenamefont {Hanikel},\ and\ \citenamefont
		{Yaghi}}]{Kalmutzki2018}%
	\BibitemOpen
	\bibfield  {author} {\bibinfo {author} {\bibfnamefont {M.~J.}\ \bibnamefont
			{Kalmutzki}}, \bibinfo {author} {\bibfnamefont {N.}~\bibnamefont {Hanikel}},\
		and\ \bibinfo {author} {\bibfnamefont {O.~M.}\ \bibnamefont {Yaghi}},\ }\href
	{https://advances.sciencemag.org/content/4/10/eaat9180} {\bibfield  {journal}
		{\bibinfo  {journal} {Sci. Adv.}\ }\textbf {\bibinfo {volume} {4}} (\bibinfo
		{year} {2018})}\BibitemShut {NoStop}%
	\bibitem [{\citenamefont {Wells}(1977)}]{Wells1977}%
	\BibitemOpen
	\bibfield  {author} {\bibinfo {author} {\bibfnamefont {A.~F.}\ \bibnamefont
			{Wells}},\ }\href@noop {} {\emph {\bibinfo {title} {Three-dimensional nets
				and polyhedra}}}\ (\bibinfo  {publisher} {Wiley, New York},\ \bibinfo {year}
	{1977})\BibitemShut {NoStop}%
	\bibitem [{\citenamefont {Ueda}\ \emph {et~al.}(1996)\citenamefont {Ueda},
		\citenamefont {Kontani}, \citenamefont {Sigrist},\ and\ \citenamefont
		{Lee}}]{Ueda1996}%
	\BibitemOpen
	\bibfield  {author} {\bibinfo {author} {\bibfnamefont {K.}~\bibnamefont
			{Ueda}}, \bibinfo {author} {\bibfnamefont {H.}~\bibnamefont {Kontani}},
		\bibinfo {author} {\bibfnamefont {M.}~\bibnamefont {Sigrist}},\ and\ \bibinfo
		{author} {\bibfnamefont {P.~A.}\ \bibnamefont {Lee}},\ }\href
	{https://doi.org/10.1103/PhysRevLett.76.1932} {\bibfield  {journal} {\bibinfo
			{journal} {Phys. Rev. Lett.}\ }\textbf {\bibinfo {volume} {76}},\ \bibinfo
		{pages} {1932} (\bibinfo {year} {1996})}\BibitemShut {NoStop}%
	\bibitem [{\citenamefont {Janani}\ \emph
		{et~al.}(2014{\natexlab{a}})\citenamefont {Janani}, \citenamefont {Merino},
		\citenamefont {McCulloch},\ and\ \citenamefont {Powell}}]{Janani2014b}%
	\BibitemOpen
	\bibfield  {author} {\bibinfo {author} {\bibfnamefont {C.}~\bibnamefont
			{Janani}}, \bibinfo {author} {\bibfnamefont {J.}~\bibnamefont {Merino}},
		\bibinfo {author} {\bibfnamefont {I.~P.}\ \bibnamefont {McCulloch}},\ and\
		\bibinfo {author} {\bibfnamefont {B.~J.}\ \bibnamefont {Powell}},\ }\href
	{http://link.aps.org/doi/10.1103/PhysRevLett.113.267204} {\bibfield
		{journal} {\bibinfo  {journal} {Phys. Rev. Lett.}\ }\textbf {\bibinfo
			{volume} {113}},\ \bibinfo {pages} {267204} (\bibinfo {year}
		{2014}{\natexlab{a}})}\BibitemShut {NoStop}%
	\bibitem [{\citenamefont {Nourse}\ \emph {et~al.}(2016)\citenamefont {Nourse},
		\citenamefont {McCulloch}, \citenamefont {Janani},\ and\ \citenamefont
		{Powell}}]{Nourse2016}%
	\BibitemOpen
	\bibfield  {author} {\bibinfo {author} {\bibfnamefont {H.~L.}\ \bibnamefont
			{Nourse}}, \bibinfo {author} {\bibfnamefont {I.~P.}\ \bibnamefont
			{McCulloch}}, \bibinfo {author} {\bibfnamefont {C.}~\bibnamefont {Janani}},\
		and\ \bibinfo {author} {\bibfnamefont {B.~J.}\ \bibnamefont {Powell}},\
	}\href {https://link.aps.org/doi/10.1103/PhysRevB.94.214418} {\bibfield
		{journal} {\bibinfo  {journal} {Phys. Rev. B}\ }\textbf {\bibinfo {volume}
			{94}},\ \bibinfo {pages} {214418} (\bibinfo {year} {2016})}\BibitemShut
	{NoStop}%
	\bibitem [{\citenamefont {Reja}\ and\ \citenamefont
		{Nishimoto}(2019)}]{Reja2019}%
	\BibitemOpen
	\bibfield  {author} {\bibinfo {author} {\bibfnamefont {S.}~\bibnamefont
			{Reja}}\ and\ \bibinfo {author} {\bibfnamefont {S.}~\bibnamefont
			{Nishimoto}},\ }\href {https://doi.org/10.1038/s41598-019-39130-4} {\bibfield
		{journal} {\bibinfo  {journal} {Sci. Rep.}\ }\textbf {\bibinfo {volume}
			{9}},\ \bibinfo {pages} {2691} (\bibinfo {year} {2019})}\BibitemShut
	{NoStop}%
	\bibitem [{\citenamefont {R{\"{u}}egg}\ \emph {et~al.}(2010)\citenamefont
		{R{\"{u}}egg}, \citenamefont {Wen},\ and\ \citenamefont {Fiete}}]{Ruegg2010}%
	\BibitemOpen
	\bibfield  {author} {\bibinfo {author} {\bibfnamefont {A.}~\bibnamefont
			{R{\"{u}}egg}}, \bibinfo {author} {\bibfnamefont {J.}~\bibnamefont {Wen}},\
		and\ \bibinfo {author} {\bibfnamefont {G.~A.}\ \bibnamefont {Fiete}},\ }\href
	{https://doi.org/10.1103/PhysRevB.81.205115} {\bibfield  {journal} {\bibinfo
			{journal} {Phys. Rev. B}\ }\textbf {\bibinfo {volume} {81}},\ \bibinfo
		{pages} {205115} (\bibinfo {year} {2010})}\BibitemShut {NoStop}%
	\bibitem [{\citenamefont {Wen}\ \emph {et~al.}(2010)\citenamefont {Wen},
		\citenamefont {R\"uegg}, \citenamefont {Wang},\ and\ \citenamefont
		{Fiete}}]{Wen2010}%
	\BibitemOpen
	\bibfield  {author} {\bibinfo {author} {\bibfnamefont {J.}~\bibnamefont
			{Wen}}, \bibinfo {author} {\bibfnamefont {A.}~\bibnamefont {R\"uegg}},
		\bibinfo {author} {\bibfnamefont {C.~C.~{\relax Joseph}.}\ \bibnamefont
			{Wang}},\ and\ \bibinfo {author} {\bibfnamefont {G.~A.}\ \bibnamefont
			{Fiete}},\ }\href {https://doi.org/10.1103/PhysRevB.82.075125} {\bibfield
		{journal} {\bibinfo  {journal} {Phys. Rev. B}\ }\textbf {\bibinfo {volume}
			{82}},\ \bibinfo {pages} {075125} (\bibinfo {year} {2010})}\BibitemShut
	{NoStop}%
	\bibitem [{\citenamefont {Chen}\ \emph {et~al.}(2018)\citenamefont {Chen},
		\citenamefont {Hui}, \citenamefont {Tewari},\ and\ \citenamefont
		{Scarola}}]{Chen2018}%
	\BibitemOpen
	\bibfield  {author} {\bibinfo {author} {\bibfnamefont {M.}~\bibnamefont
			{Chen}}, \bibinfo {author} {\bibfnamefont {H.-Y.}\ \bibnamefont {Hui}},
		\bibinfo {author} {\bibfnamefont {S.}~\bibnamefont {Tewari}},\ and\ \bibinfo
		{author} {\bibfnamefont {V.~W.}\ \bibnamefont {Scarola}},\ }\href
	{https://doi.org/10.1103/PhysRevB.97.035114} {\bibfield  {journal} {\bibinfo
			{journal} {Phys. Rev. B}\ }\textbf {\bibinfo {volume} {97}},\ \bibinfo
		{pages} {035114} (\bibinfo {year} {2018})}\BibitemShut {NoStop}%
	\bibitem [{\citenamefont {L\'opez}\ and\ \citenamefont
		{Merino}(2019)}]{Lopez2019}%
	\BibitemOpen
	\bibfield  {author} {\bibinfo {author} {\bibfnamefont {M.~F.}\ \bibnamefont
			{L\'opez}}\ and\ \bibinfo {author} {\bibfnamefont {J.}~\bibnamefont
			{Merino}},\ }\href {https://doi.org/10.1103/PhysRevB.100.075154} {\bibfield
		{journal} {\bibinfo  {journal} {Phys. Rev. B}\ }\textbf {\bibinfo {volume}
			{100}},\ \bibinfo {pages} {075154} (\bibinfo {year} {2019})}\BibitemShut
	{NoStop}%
	\bibitem [{\citenamefont {Yao}\ \emph {et~al.}(2007)\citenamefont {Yao},
		\citenamefont {Tsai},\ and\ \citenamefont {Kivelson}}]{Yao2007}%
	\BibitemOpen
	\bibfield  {author} {\bibinfo {author} {\bibfnamefont {H.}~\bibnamefont
			{Yao}}, \bibinfo {author} {\bibfnamefont {W.-F.}\ \bibnamefont {Tsai}},\ and\
		\bibinfo {author} {\bibfnamefont {S.~A.}\ \bibnamefont {Kivelson}},\ }\href
	{https://doi.org/10.1103/PhysRevB.76.161104} {\bibfield  {journal} {\bibinfo
			{journal} {Phys. Rev. B}\ }\textbf {\bibinfo {volume} {76}},\ \bibinfo
		{pages} {161104(R)} (\bibinfo {year} {2007})}\BibitemShut {NoStop}%
	\bibitem [{\citenamefont {Sur}\ \emph {et~al.}(2018)\citenamefont {Sur},
		\citenamefont {Gong}, \citenamefont {Yang},\ and\ \citenamefont
		{Vafek}}]{Sur2018}%
	\BibitemOpen
	\bibfield  {author} {\bibinfo {author} {\bibfnamefont {S.}~\bibnamefont
			{Sur}}, \bibinfo {author} {\bibfnamefont {S.-S.}\ \bibnamefont {Gong}},
		\bibinfo {author} {\bibfnamefont {K.}~\bibnamefont {Yang}},\ and\ \bibinfo
		{author} {\bibfnamefont {O.}~\bibnamefont {Vafek}},\ }\href
	{https://doi.org/10.1103/PhysRevB.98.125144} {\bibfield  {journal} {\bibinfo
			{journal} {Phys. Rev. B}\ }\textbf {\bibinfo {volume} {98}},\ \bibinfo
		{pages} {125144} (\bibinfo {year} {2018})}\BibitemShut {NoStop}%
	\bibitem [{\citenamefont {Dagotto}(2013)}]{Dagotto2013}%
	\BibitemOpen
	\bibfield  {author} {\bibinfo {author} {\bibfnamefont {E.}~\bibnamefont
			{Dagotto}},\ }\href {https://doi.org/10.1103/RevModPhys.85.849} {\bibfield
		{journal} {\bibinfo  {journal} {Rev. Mod. Phys.}\ }\textbf {\bibinfo {volume}
			{85}},\ \bibinfo {pages} {849} (\bibinfo {year} {2013})}\BibitemShut
	{NoStop}%
	\bibitem [{\citenamefont {Yanagi}\ and\ \citenamefont
		{Ueda}(2014)}]{Yanagi2014}%
	\BibitemOpen
	\bibfield  {author} {\bibinfo {author} {\bibfnamefont {Y.}~\bibnamefont
			{Yanagi}}\ and\ \bibinfo {author} {\bibfnamefont {K.}~\bibnamefont {Ueda}},\
	}\href {https://doi.org/10.1103/PhysRevB.90.085113} {\bibfield  {journal}
		{\bibinfo  {journal} {Phys. Rev. B}\ }\textbf {\bibinfo {volume} {90}},\
		\bibinfo {pages} {085113} (\bibinfo {year} {2014})}\BibitemShut {NoStop}%
	\bibitem [{\citenamefont {Khatami}\ \emph {et~al.}(2014)\citenamefont
		{Khatami}, \citenamefont {Singh}, \citenamefont {Pickett},\ and\
		\citenamefont {Scalettar}}]{Khatami2014}%
	\BibitemOpen
	\bibfield  {author} {\bibinfo {author} {\bibfnamefont {E.}~\bibnamefont
			{Khatami}}, \bibinfo {author} {\bibfnamefont {R.~R.~P.}\ \bibnamefont
			{Singh}}, \bibinfo {author} {\bibfnamefont {W.~E.}\ \bibnamefont {Pickett}},\
		and\ \bibinfo {author} {\bibfnamefont {R.~T.}\ \bibnamefont {Scalettar}},\
	}\href {https://doi.org/10.1103/PhysRevLett.113.106402} {\bibfield  {journal}
		{\bibinfo  {journal} {Phys. Rev. Lett.}\ }\textbf {\bibinfo {volume} {113}},\
		\bibinfo {pages} {106402} (\bibinfo {year} {2014})}\BibitemShut {NoStop}%
	\bibitem [{\citenamefont {Feng}\ \emph {et~al.}(2020)\citenamefont {Feng},
		\citenamefont {Guo},\ and\ \citenamefont {Scalettar}}]{Feng2019}%
	\BibitemOpen
	\bibfield  {author} {\bibinfo {author} {\bibfnamefont {C.}~\bibnamefont
			{Feng}}, \bibinfo {author} {\bibfnamefont {H.}~\bibnamefont {Guo}},\ and\
		\bibinfo {author} {\bibfnamefont {R.~T.}\ \bibnamefont {Scalettar}},\ }\href
	{https://doi.org/10.1103/PhysRevB.101.205103} {\bibfield  {journal} {\bibinfo
			{journal} {Phys. Rev. B}\ }\textbf {\bibinfo {volume} {101}},\ \bibinfo
		{pages} {205103} (\bibinfo {year} {2020})}\BibitemShut {NoStop}%
	\bibitem [{\citenamefont {Lin}\ \emph {et~al.}(2014)\citenamefont {Lin},
		\citenamefont {Chen}, \citenamefont {Liu}, \citenamefont {Tao},\ and\
		\citenamefont {Liu}}]{Lin2014}%
	\BibitemOpen
	\bibfield  {author} {\bibinfo {author} {\bibfnamefont {H.-F.}\ \bibnamefont
			{Lin}}, \bibinfo {author} {\bibfnamefont {Y.-H.}\ \bibnamefont {Chen}},
		\bibinfo {author} {\bibfnamefont {H.-D.}\ \bibnamefont {Liu}}, \bibinfo
		{author} {\bibfnamefont {H.-S.}\ \bibnamefont {Tao}},\ and\ \bibinfo {author}
		{\bibfnamefont {W.-M.}\ \bibnamefont {Liu}},\ }\href
	{https://doi.org/10.1103/PhysRevA.90.053627} {\bibfield  {journal} {\bibinfo
			{journal} {Phys. Rev. A}\ }\textbf {\bibinfo {volume} {90}},\ \bibinfo
		{pages} {053627} (\bibinfo {year} {2014})}\BibitemShut {NoStop}%
	\bibitem [{\citenamefont {Richter}\ \emph {et~al.}(2004)\citenamefont
		{Richter}, \citenamefont {Schulenburg}, \citenamefont {Honecker},\ and\
		\citenamefont {Schmalfu{\ss}}}]{Richter2004}%
	\BibitemOpen
	\bibfield  {author} {\bibinfo {author} {\bibfnamefont {J.}~\bibnamefont
			{Richter}}, \bibinfo {author} {\bibfnamefont {J.}~\bibnamefont
			{Schulenburg}}, \bibinfo {author} {\bibfnamefont {A.}~\bibnamefont
			{Honecker}},\ and\ \bibinfo {author} {\bibfnamefont {D.}~\bibnamefont
			{Schmalfu{\ss}}},\ }\href
	{https://link.aps.org/doi/10.1103/PhysRevB.70.174454} {\bibfield  {journal}
		{\bibinfo  {journal} {Phys. Rev. B}\ }\textbf {\bibinfo {volume} {70}},\
		\bibinfo {pages} {174454} (\bibinfo {year} {2004})}\BibitemShut {NoStop}%
	\bibitem [{\citenamefont {Misguich}\ and\ \citenamefont
		{Sindzingre}(2007)}]{Misguich2007}%
	\BibitemOpen
	\bibfield  {author} {\bibinfo {author} {\bibfnamefont {G.}~\bibnamefont
			{Misguich}}\ and\ \bibinfo {author} {\bibfnamefont {P.}~\bibnamefont
			{Sindzingre}},\ }\href {http://dx.doi.org/10.1088/0953-8984/19/14/145202}
	{\bibfield  {journal} {\bibinfo  {journal} {J. Phys.: Condens. Matter}\
		}\textbf {\bibinfo {volume} {19}},\ \bibinfo {pages} {145202} (\bibinfo
		{year} {2007})}\BibitemShut {NoStop}%
	\bibitem [{\citenamefont {Yang}\ \emph {et~al.}(2010)\citenamefont {Yang},
		\citenamefont {Paramekanti},\ and\ \citenamefont {Kim}}]{Yang2010}%
	\BibitemOpen
	\bibfield  {author} {\bibinfo {author} {\bibfnamefont {B.-J.}\ \bibnamefont
			{Yang}}, \bibinfo {author} {\bibfnamefont {A.}~\bibnamefont {Paramekanti}},\
		and\ \bibinfo {author} {\bibfnamefont {Y.~B.}\ \bibnamefont {Kim}},\ }\href
	{https://link.aps.org/doi/10.1103/PhysRevB.81.134418} {\bibfield  {journal}
		{\bibinfo  {journal} {Phys. Rev. B}\ }\textbf {\bibinfo {volume} {81}},\
		\bibinfo {pages} {134418} (\bibinfo {year} {2010})}\BibitemShut {NoStop}%
	\bibitem [{\citenamefont {Jahromi}\ and\ \citenamefont
		{Or{\'{u}}s}(2018)}]{Jahromi2018}%
	\BibitemOpen
	\bibfield  {author} {\bibinfo {author} {\bibfnamefont {S.~S.}\ \bibnamefont
			{Jahromi}}\ and\ \bibinfo {author} {\bibfnamefont {R.}~\bibnamefont
			{Or{\'{u}}s}},\ }\href {https://link.aps.org/doi/10.1103/PhysRevB.98.155108}
	{\bibfield  {journal} {\bibinfo  {journal} {Phys. Rev. B}\ }\textbf {\bibinfo
			{volume} {98}},\ \bibinfo {pages} {155108} (\bibinfo {year}
		{2018})}\BibitemShut {NoStop}%
	\bibitem [{\citenamefont {Kitaev}(2006)}]{Kitaev2006}%
	\BibitemOpen
	\bibfield  {author} {\bibinfo {author} {\bibfnamefont {A.}~\bibnamefont
			{Kitaev}},\ }\href
	{https://doi.org/https://doi.org/10.1016/j.aop.2005.10.005} {\bibfield
		{journal} {\bibinfo  {journal} {Ann. Phys.}\ }\textbf {\bibinfo {volume}
			{321}},\ \bibinfo {pages} {2} (\bibinfo {year} {2006})}\BibitemShut {NoStop}%
	\bibitem [{\citenamefont {Yao}\ and\ \citenamefont
		{Kivelson}(2007)}]{Yao2007b}%
	\BibitemOpen
	\bibfield  {author} {\bibinfo {author} {\bibfnamefont {H.}~\bibnamefont
			{Yao}}\ and\ \bibinfo {author} {\bibfnamefont {S.~A.}\ \bibnamefont
			{Kivelson}},\ }\href {https://doi.org/10.1103/PhysRevLett.99.247203}
	{\bibfield  {journal} {\bibinfo  {journal} {Phys. Rev. Lett.}\ }\textbf
		{\bibinfo {volume} {99}},\ \bibinfo {pages} {247203} (\bibinfo {year}
		{2007})}\BibitemShut {NoStop}%
	\bibitem [{\citenamefont {Dusuel}\ \emph {et~al.}(2008)\citenamefont {Dusuel},
		\citenamefont {Schmidt}, \citenamefont {Vidal},\ and\ \citenamefont
		{Zaffino}}]{Dusuel2008}%
	\BibitemOpen
	\bibfield  {author} {\bibinfo {author} {\bibfnamefont {S.}~\bibnamefont
			{Dusuel}}, \bibinfo {author} {\bibfnamefont {K.~P.}\ \bibnamefont {Schmidt}},
		\bibinfo {author} {\bibfnamefont {J.}~\bibnamefont {Vidal}},\ and\ \bibinfo
		{author} {\bibfnamefont {R.~L.}\ \bibnamefont {Zaffino}},\ }\href
	{https://doi.org/10.1103/PhysRevB.78.125102} {\bibfield  {journal} {\bibinfo
			{journal} {Phys. Rev. B}\ }\textbf {\bibinfo {volume} {78}},\ \bibinfo
		{pages} {125102} (\bibinfo {year} {2008})}\BibitemShut {NoStop}%
	\bibitem [{\citenamefont {Khosla}\ \emph {et~al.}(2017)\citenamefont {Khosla},
		\citenamefont {Jacko}, \citenamefont {Merino},\ and\ \citenamefont
		{Powell}}]{Khosla2017}%
	\BibitemOpen
	\bibfield  {author} {\bibinfo {author} {\bibfnamefont {A.~L.}\ \bibnamefont
			{Khosla}}, \bibinfo {author} {\bibfnamefont {A.~C.}\ \bibnamefont {Jacko}},
		\bibinfo {author} {\bibfnamefont {J.}~\bibnamefont {Merino}},\ and\ \bibinfo
		{author} {\bibfnamefont {B.~J.}\ \bibnamefont {Powell}},\ }\href
	{https://doi.org/10.1103/PhysRevB.95.115109} {\bibfield  {journal} {\bibinfo
			{journal} {Phys. Rev. B}\ }\textbf {\bibinfo {volume} {95}},\ \bibinfo
		{pages} {115109} (\bibinfo {year} {2017})}\BibitemShut {NoStop}%
	\bibitem [{\citenamefont {Powell}\ \emph {et~al.}(2017)\citenamefont {Powell},
		\citenamefont {Merino}, \citenamefont {Khosla},\ and\ \citenamefont
		{Jacko}}]{Powell2017}%
	\BibitemOpen
	\bibfield  {author} {\bibinfo {author} {\bibfnamefont {B.~J.}\ \bibnamefont
			{Powell}}, \bibinfo {author} {\bibfnamefont {J.}~\bibnamefont {Merino}},
		\bibinfo {author} {\bibfnamefont {A.~L.}\ \bibnamefont {Khosla}},\ and\
		\bibinfo {author} {\bibfnamefont {A.~C.}\ \bibnamefont {Jacko}},\ }\href
	{https://doi.org/10.1103/PhysRevB.95.094432} {\bibfield  {journal} {\bibinfo
			{journal} {Phys. Rev. B}\ }\textbf {\bibinfo {volume} {95}},\ \bibinfo
		{pages} {094432} (\bibinfo {year} {2017})}\BibitemShut {NoStop}%
	\bibitem [{\citenamefont {Si}\ \emph {et~al.}(2016)\citenamefont {Si},
		\citenamefont {Yu},\ and\ \citenamefont {Abrahams}}]{Si2016}%
	\BibitemOpen
	\bibfield  {author} {\bibinfo {author} {\bibfnamefont {Q.}~\bibnamefont
			{Si}}, \bibinfo {author} {\bibfnamefont {R.}~\bibnamefont {Yu}},\ and\
		\bibinfo {author} {\bibfnamefont {E.}~\bibnamefont {Abrahams}},\ }\href
	{https://doi.org/10.1038/natrevmats.2016.17} {\bibfield  {journal} {\bibinfo
			{journal} {Nat. Rev. Mater.}\ }\textbf {\bibinfo {volume} {1}},\ \bibinfo
		{pages} {16017} (\bibinfo {year} {2016})}\BibitemShut {NoStop}%
	\bibitem [{\citenamefont {Kotliar}\ and\ \citenamefont
		{Ruckenstein}(1986)}]{Kotliar1986}%
	\BibitemOpen
	\bibfield  {author} {\bibinfo {author} {\bibfnamefont {G.}~\bibnamefont
			{Kotliar}}\ and\ \bibinfo {author} {\bibfnamefont {A.~E.}\ \bibnamefont
			{Ruckenstein}},\ }\href {http://link.aps.org/doi/10.1103/PhysRevLett.57.1362}
	{\bibfield  {journal} {\bibinfo  {journal} {Phys. Rev. Lett.}\ }\textbf
		{\bibinfo {volume} {57}},\ \bibinfo {pages} {1362} (\bibinfo {year}
		{1986})}\BibitemShut {NoStop}%
	\bibitem [{\citenamefont {Lechermann}\ \emph {et~al.}(2007)\citenamefont
		{Lechermann}, \citenamefont {Georges}, \citenamefont {Kotliar},\ and\
		\citenamefont {Parcollet}}]{Lechermann2007}%
	\BibitemOpen
	\bibfield  {author} {\bibinfo {author} {\bibfnamefont {F.}~\bibnamefont
			{Lechermann}}, \bibinfo {author} {\bibfnamefont {A.}~\bibnamefont {Georges}},
		\bibinfo {author} {\bibfnamefont {G.}~\bibnamefont {Kotliar}},\ and\ \bibinfo
		{author} {\bibfnamefont {O.}~\bibnamefont {Parcollet}},\ }\href
	{http://link.aps.org/doi/10.1103/PhysRevB.76.155102} {\bibfield  {journal}
		{\bibinfo  {journal} {Phys. Rev. B}\ }\textbf {\bibinfo {volume} {76}},\
		\bibinfo {pages} {155102} (\bibinfo {year} {2007})}\BibitemShut {NoStop}%
	\bibitem [{\citenamefont {Lanat{\`{a}}}\ \emph {et~al.}(2015)\citenamefont
		{Lanat{\`{a}}}, \citenamefont {Yao}, \citenamefont {Wang}, \citenamefont
		{Ho},\ and\ \citenamefont {Kotliar}}]{Lanata2015}%
	\BibitemOpen
	\bibfield  {author} {\bibinfo {author} {\bibfnamefont {N.}~\bibnamefont
			{Lanat{\`{a}}}}, \bibinfo {author} {\bibfnamefont {Y.}~\bibnamefont {Yao}},
		\bibinfo {author} {\bibfnamefont {C.-Z.}\ \bibnamefont {Wang}}, \bibinfo
		{author} {\bibfnamefont {K.-M.}\ \bibnamefont {Ho}},\ and\ \bibinfo {author}
		{\bibfnamefont {G.}~\bibnamefont {Kotliar}},\ }\href
	{http://link.aps.org/doi/10.1103/PhysRevX.5.011008} {\bibfield  {journal}
		{\bibinfo  {journal} {Phys. Rev. X}\ }\textbf {\bibinfo {volume} {5}},\
		\bibinfo {pages} {11008} (\bibinfo {year} {2015})}\BibitemShut {NoStop}%
	\bibitem [{\citenamefont {Lanat{\`{a}}}\ \emph {et~al.}(2017)\citenamefont
		{Lanat{\`{a}}}, \citenamefont {Yao}, \citenamefont {Deng}, \citenamefont
		{Dobrosavljevi{\'{c}}},\ and\ \citenamefont {Kotliar}}]{Lanata2017}%
	\BibitemOpen
	\bibfield  {author} {\bibinfo {author} {\bibfnamefont {N.}~\bibnamefont
			{Lanat{\`{a}}}}, \bibinfo {author} {\bibfnamefont {Y.}~\bibnamefont {Yao}},
		\bibinfo {author} {\bibfnamefont {X.}~\bibnamefont {Deng}}, \bibinfo {author}
		{\bibfnamefont {V.}~\bibnamefont {Dobrosavljevi{\'{c}}}},\ and\ \bibinfo
		{author} {\bibfnamefont {G.}~\bibnamefont {Kotliar}},\ }\href
	{https://link.aps.org/doi/10.1103/PhysRevLett.118.126401} {\bibfield
		{journal} {\bibinfo  {journal} {Phys. Rev. Lett.}\ }\textbf {\bibinfo
			{volume} {118}},\ \bibinfo {pages} {126401} (\bibinfo {year}
		{2017})}\BibitemShut {NoStop}%
	\bibitem [{sup()}]{sup}%
	\BibitemOpen
	\href@noop {} {}\bibinfo {note} {See Supplemental Material at [URL will be
		inserted by publisher], which includes Refs.~\cite{Isidori2009,Bunemann2003,Lanata2017b,Parcollet2015,Seth2016,Blochl1994,TRIQS-ARPACK}, for technical details on RISB, the relation between
		the quasiparticle weight of the bands and the molecular orbitals, the
		explicit form of $\hat{H}$ in the dimer and trimer basis, and the magnetic
		phase diagram in the honeycomb-like regime.}\BibitemShut {Stop}%
	\bibitem [{\citenamefont {Isidori}\ and\ \citenamefont
		{Capone}(2009)}]{Isidori2009}%
	\BibitemOpen
	\bibfield  {author} {\bibinfo {author} {\bibfnamefont {A.}~\bibnamefont
			{Isidori}}\ and\ \bibinfo {author} {\bibfnamefont {M.}~\bibnamefont
			{Capone}},\ }\href {https://doi.org/10.1103/PhysRevB.80.115120} {\bibfield
		{journal} {\bibinfo  {journal} {Phys. Rev. B}\ }\textbf {\bibinfo {volume}
			{80}},\ \bibinfo {pages} {115120} (\bibinfo {year} {2009})}\BibitemShut
	{NoStop}%
	\bibitem [{\citenamefont {B\"unemann}\ \emph {et~al.}(2003)\citenamefont
		{B\"unemann}, \citenamefont {Gebhard},\ and\ \citenamefont
		{Thul}}]{Bunemann2003}%
	\BibitemOpen
	\bibfield  {author} {\bibinfo {author} {\bibfnamefont {J.}~\bibnamefont
			{B\"unemann}}, \bibinfo {author} {\bibfnamefont {F.}~\bibnamefont {Gebhard}},\
		and\ \bibinfo {author} {\bibfnamefont {R.}~\bibnamefont {Thul}},\ }\href
	{https://doi.org/10.1103/PhysRevB.67.075103} {\bibfield  {journal} {\bibinfo
			{journal} {Phys. Rev. B}\ }\textbf {\bibinfo {volume} {67}},\ \bibinfo
		{pages} {75103} (\bibinfo {year} {2003})}\BibitemShut {NoStop}%
	\bibitem [{\citenamefont {Lanat{\`{a}}}\ \emph {et~al.}(2017)\citenamefont
		{Lanat{\`{a}}}, \citenamefont {Lee}, \citenamefont {Yao},\ and\ \citenamefont
		{Dobrosavljevi{\'{c}}}}]{Lanata2017b}%
	\BibitemOpen
	\bibfield  {author} {\bibinfo {author} {\bibfnamefont {N.}~\bibnamefont
			{Lanat{\`{a}}}}, \bibinfo {author} {\bibfnamefont {T.-H.}\ \bibnamefont {Lee}},
		\bibinfo {author} {\bibfnamefont {Y.-X.}\ \bibnamefont {Yao}},\ and\ \bibinfo
		{author} {\bibfnamefont {V.}~\bibnamefont {Dobrosavljevi{\'{c}}}},\ }\href
	{https://doi.org/10.1103/PhysRevB.96.195126} {\bibfield  {journal} {\bibinfo
			{journal} {Phys. Rev. B}\ }\textbf {\bibinfo {volume} {96}},\ \bibinfo
		{pages} {195126} (\bibinfo {year} {2017})}\BibitemShut {NoStop}%
	\bibitem [{\citenamefont {Parcollet}\ \emph {et~al.}(2015)\citenamefont
		{Parcollet}, \citenamefont {Ferrero}, \citenamefont {Ayral}, \citenamefont
		{Hafermann}, \citenamefont {Krivenko}, \citenamefont {Messio},\ and\
		\citenamefont {Seth}}]{Parcollet2015}%
	\BibitemOpen
	\bibfield  {author} {\bibinfo {author} {\bibfnamefont {O.}~\bibnamefont
			{Parcollet}}, \bibinfo {author} {\bibfnamefont {M.}~\bibnamefont {Ferrero}},
		\bibinfo {author} {\bibfnamefont {T.}~\bibnamefont {Ayral}}, \bibinfo
		{author} {\bibfnamefont {H.}~\bibnamefont {Hafermann}}, \bibinfo {author}
		{\bibfnamefont {I.}~\bibnamefont {Krivenko}}, \bibinfo {author}
		{\bibfnamefont {L.}~\bibnamefont {Messio}},\ and\ \bibinfo {author}
		{\bibfnamefont {P.}~\bibnamefont {Seth}},\ }\href
	{http://www.sciencedirect.com/science/article/pii/S0010465515001666}
	{\bibfield  {journal} {\bibinfo  {journal} {Comput. Phys. Commun.}\ }\textbf
		{\bibinfo {volume} {196}},\ \bibinfo {pages} {398} (\bibinfo {year}
		{2015})},\ \bibinfo {note} {version 1.4}\BibitemShut {NoStop}%
	\bibitem [{\citenamefont {Seth}\ \emph {et~al.}(2016)\citenamefont {Seth},
		\citenamefont {Krivenko}, \citenamefont {Ferrero},\ and\ \citenamefont
		{Parcollet}}]{Seth2016}%
	\BibitemOpen
	\bibfield  {author} {\bibinfo {author} {\bibfnamefont {P.}~\bibnamefont
			{Seth}}, \bibinfo {author} {\bibfnamefont {I.}~\bibnamefont {Krivenko}},
		\bibinfo {author} {\bibfnamefont {M.}~\bibnamefont {Ferrero}},\ and\ \bibinfo
		{author} {\bibfnamefont {O.}~\bibnamefont {Parcollet}},\ }\href
	{http://www.sciencedirect.com/science/article/pii/S001046551500404X}
	{\bibfield  {journal} {\bibinfo  {journal} {Comput. Phys. Commun.}\ }\textbf
		{\bibinfo {volume} {200}},\ \bibinfo {pages} {274} (\bibinfo {year}
		{2016})}\BibitemShut {NoStop}%
	\bibitem [{\citenamefont {Bl{\"{o}}chl}\ \emph {et~al.}(1994)\citenamefont
		{Bl{\"{o}}chl}, \citenamefont {Jepsen},\ and\ \citenamefont
		{Andersen}}]{Blochl1994}%
	\BibitemOpen
	\bibfield  {author} {\bibinfo {author} {\bibfnamefont {P.~E.}\ \bibnamefont
			{Bl{\"{o}}chl}}, \bibinfo {author} {\bibfnamefont {O.}~\bibnamefont
			{Jepsen}},\ and\ \bibinfo {author} {\bibfnamefont {O.~K.}\ \bibnamefont
			{Andersen}},\ }\href {http://link.aps.org/doi/10.1103/PhysRevB.49.16223}
	{\bibfield  {journal} {\bibinfo  {journal} {Phys. Rev. B}\ }\textbf {\bibinfo
			{volume} {49}},\ \bibinfo {pages} {16223} (\bibinfo {year}
		{1994})}\BibitemShut {NoStop}%
	\bibitem [{\citenamefont {Krivenko}(2020)}]{TRIQS-ARPACK}%
	\BibitemOpen
	\bibfield  {author} {\bibinfo {author} {\bibfnamefont {I.}~\bibnamefont
			{Krivenko}},\ }\href {https://doi.org/10.5281/zenodo.3930202} {\bibinfo
		{title} {{ezARPACK - a C++ ARPACK-NG wrapper compatible with multiple
				matrix/vector algebra libraries: Release 0.9}}} (\bibinfo {year}
	{2020})\BibitemShut {NoStop}%
	\bibitem [{\citenamefont {B\"unemann}\ and\ \citenamefont
		{Gebhard}(2007)}]{Bunemann2007}%
	\BibitemOpen
	\bibfield  {author} {\bibinfo {author} {\bibfnamefont {J.}~\bibnamefont
			{B\"unemann}}\ and\ \bibinfo {author} {\bibfnamefont {F.}~\bibnamefont
			{Gebhard}},\ }\href {https://link.aps.org/doi/10.1103/PhysRevB.76.193104}
	{\bibfield  {journal} {\bibinfo  {journal} {Phys. Rev. B}\ }\textbf {\bibinfo
			{volume} {76}},\ \bibinfo {pages} {193104} (\bibinfo {year}
		{2007})}\BibitemShut {NoStop}%
	\bibitem [{\citenamefont {Burdin}\ \emph {et~al.}(2000)\citenamefont {Burdin},
		\citenamefont {Georges},\ and\ \citenamefont {Grempel}}]{Burdin2000}%
	\BibitemOpen
	\bibfield  {author} {\bibinfo {author} {\bibfnamefont {S.}~\bibnamefont
			{Burdin}}, \bibinfo {author} {\bibfnamefont {A.}~\bibnamefont {Georges}},\
		and\ \bibinfo {author} {\bibfnamefont {D.~R.}\ \bibnamefont {Grempel}},\
	}\href {https://doi.org/10.1103/PhysRevLett.85.1048} {\bibfield  {journal}
		{\bibinfo  {journal} {Phys. Rev. Lett.}\ }\textbf {\bibinfo {volume} {85}},\
		\bibinfo {pages} {1048} (\bibinfo {year} {2000})}\BibitemShut {NoStop}%
	\bibitem [{\citenamefont {Zhu}\ \emph {et~al.}(2004)\citenamefont {Zhu},
		\citenamefont {Kirchner}, \citenamefont {Si},\ and\ \citenamefont
		{Georges}}]{Zhu2004}%
	\BibitemOpen
	\bibfield  {author} {\bibinfo {author} {\bibfnamefont {L.}~\bibnamefont
			{Zhu}}, \bibinfo {author} {\bibfnamefont {S.}~\bibnamefont {Kirchner}},
		\bibinfo {author} {\bibfnamefont {Q.}~\bibnamefont {Si}},\ and\ \bibinfo
		{author} {\bibfnamefont {A.}~\bibnamefont {Georges}},\ }\href
	{https://doi.org/10.1103/PhysRevLett.93.267201} {\bibfield  {journal}
		{\bibinfo  {journal} {Phys. Rev. Lett.}\ }\textbf {\bibinfo {volume} {93}},\
		\bibinfo {pages} {267201} (\bibinfo {year} {2004})}\BibitemShut {NoStop}%
	\bibitem [{\citenamefont {Lechermann}(2009)}]{Lechermann2009}%
	\BibitemOpen
	\bibfield  {author} {\bibinfo {author} {\bibfnamefont {F.}~\bibnamefont
			{Lechermann}},\ }\href {https://doi.org/10.1103/PhysRevLett.102.046403}
	{\bibfield  {journal} {\bibinfo  {journal} {Phys. Rev. Lett.}\ }\textbf
		{\bibinfo {volume} {102}},\ \bibinfo {pages} {46403} (\bibinfo {year}
		{2009})}\BibitemShut {NoStop}%
	\bibitem [{\citenamefont {Lu}\ \emph {et~al.}(2013)\citenamefont {Lu},
		\citenamefont {Zhao}, \citenamefont {Weng}, \citenamefont {Fang},\ and\
		\citenamefont {Dai}}]{Lu2013}%
	\BibitemOpen
	\bibfield  {author} {\bibinfo {author} {\bibfnamefont {F.}~\bibnamefont
			{Lu}}, \bibinfo {author} {\bibfnamefont {J. Z.}~\bibnamefont {Zhao}}, \bibinfo
		{author} {\bibfnamefont {H.}~\bibnamefont {Weng}}, \bibinfo {author}
		{\bibfnamefont {Z.}~\bibnamefont {Fang}},\ and\ \bibinfo {author}
		{\bibfnamefont {X.}~\bibnamefont {Dai}},\ }\href
	{https://doi.org/10.1103/PhysRevLett.110.096401} {\bibfield  {journal}
		{\bibinfo  {journal} {Phys. Rev. Lett.}\ }\textbf {\bibinfo {volume} {110}},\
		\bibinfo {pages} {96401} (\bibinfo {year} {2013})}\BibitemShut {NoStop}%
	\bibitem [{\citenamefont {Isidori}\ \emph {et~al.}(2019)\citenamefont
		{Isidori}, \citenamefont {Berovi{\'c}}, \citenamefont {Fanfarillo},
		\citenamefont {{de' Medici}}, \citenamefont {Fabrizio},\ and\ \citenamefont
		{Capone}}]{Isidori2019}%
	\BibitemOpen
	\bibfield  {author} {\bibinfo {author} {\bibfnamefont {A.}~\bibnamefont
			{Isidori}}, \bibinfo {author} {\bibfnamefont {M.}~\bibnamefont
			{Berovi{\'c}}}, \bibinfo {author} {\bibfnamefont {L.}~\bibnamefont
			{Fanfarillo}}, \bibinfo {author} {\bibfnamefont {L.}~\bibnamefont {{de'
					Medici}}}, \bibinfo {author} {\bibfnamefont {M.}~\bibnamefont {Fabrizio}},\
		and\ \bibinfo {author} {\bibfnamefont {M.}~\bibnamefont {Capone}},\ }\href
	{https://doi.org/10.1103/PhysRevLett.122.186401} {\bibfield  {journal}
		{\bibinfo  {journal} {Phys. Rev. Lett.}\ }\textbf {\bibinfo {volume} {122}},\
		\bibinfo {pages} {186401} (\bibinfo {year} {2019})}\BibitemShut {NoStop}%
	\bibitem [{\citenamefont {Brinkman}\ and\ \citenamefont
		{Rice}(1970)}]{Brinkman1970}%
	\BibitemOpen
	\bibfield  {author} {\bibinfo {author} {\bibfnamefont {W.~F.}\ \bibnamefont
			{Brinkman}}\ and\ \bibinfo {author} {\bibfnamefont {T.~M.}\ \bibnamefont
			{Rice}},\ }\href {https://doi.org/10.1103/PhysRevB.2.4302} {\bibfield
		{journal} {\bibinfo  {journal} {Phys. Rev. B}\ }\textbf {\bibinfo {volume}
			{2}},\ \bibinfo {pages} {4302} (\bibinfo {year} {1970})}\BibitemShut
	{NoStop}%
	\bibitem [{\citenamefont {Merino}\ \emph {et~al.}(2006)\citenamefont {Merino},
		\citenamefont {Powell},\ and\ \citenamefont {McKenzie}}]{Merino2006}%
	\BibitemOpen
	\bibfield  {author} {\bibinfo {author} {\bibfnamefont {J.}~\bibnamefont
			{Merino}}, \bibinfo {author} {\bibfnamefont {B.~J.}\ \bibnamefont {Powell}},\
		and\ \bibinfo {author} {\bibfnamefont {R.~H.}\ \bibnamefont {McKenzie}},\
	}\href {https://link.aps.org/doi/10.1103/PhysRevB.73.235107} {\bibfield
		{journal} {\bibinfo  {journal} {Phys. Rev. B}\ }\textbf {\bibinfo {volume}
			{73}},\ \bibinfo {pages} {235107} (\bibinfo {year} {2006})}\BibitemShut
	{NoStop}%
	\bibitem [{\citenamefont {Ohashi}\ \emph {et~al.}(2006)\citenamefont {Ohashi},
		\citenamefont {Kawakami},\ and\ \citenamefont {Tsunetsugu}}]{Ohashi2006}%
	\BibitemOpen
	\bibfield  {author} {\bibinfo {author} {\bibfnamefont {T.}~\bibnamefont
			{Ohashi}}, \bibinfo {author} {\bibfnamefont {N.}~\bibnamefont {Kawakami}},\
		and\ \bibinfo {author} {\bibfnamefont {H.}~\bibnamefont {Tsunetsugu}},\
	}\href {https://doi.org/10.1103/PhysRevLett.97.066401} {\bibfield  {journal}
		{\bibinfo  {journal} {Phys. Rev. Lett.}\ }\textbf {\bibinfo {volume} {97}},\
		\bibinfo {pages} {66401} (\bibinfo {year} {2006})}\BibitemShut {NoStop}%
	\bibitem [{\citenamefont {Broholm}\ \emph {et~al.}(2020)\citenamefont
		{Broholm}, \citenamefont {Cava}, \citenamefont {Kivelson}, \citenamefont
		{Nocera}, \citenamefont {Norman},\ and\ \citenamefont
		{Senthil}}]{Broholm2020}%
	\BibitemOpen
	\bibfield  {author} {\bibinfo {author} {\bibfnamefont {C.}~\bibnamefont
			{Broholm}}, \bibinfo {author} {\bibfnamefont {R.~J.}\ \bibnamefont {Cava}},
		\bibinfo {author} {\bibfnamefont {S.~A.}\ \bibnamefont {Kivelson}}, \bibinfo
		{author} {\bibfnamefont {D.~G.}\ \bibnamefont {Nocera}}, \bibinfo {author}
		{\bibfnamefont {M.~R.}\ \bibnamefont {Norman}},\ and\ \bibinfo {author}
		{\bibfnamefont {T.}~\bibnamefont {Senthil}},\ }\href
	{https://science.sciencemag.org/content/367/6475/eaay0668} {\bibfield
		{journal} {\bibinfo  {journal} {Science}\ }\textbf {\bibinfo {volume} {367}}, {\bibinfo {pages} {eaay0668}}
		(\bibinfo {year} {2020})}\BibitemShut {NoStop}%
	\bibitem [{\citenamefont {Kanoda}(1997)}]{Kanoda1997}%
	\BibitemOpen
	\bibfield  {author} {\bibinfo {author} {\bibfnamefont {K.}~\bibnamefont
			{Kanoda}},\ }\href
	{http://www.sciencedirect.com/science/article/pii/S0921453497002669}
	{\bibfield  {journal} {\bibinfo  {journal} {Phys. C Supercond.}\ }\textbf
		{\bibinfo {volume} {282-287}},\ \bibinfo {pages} {299} (\bibinfo {year}
		{1997})}\BibitemShut {NoStop}%
	%
	\bibitem [{\citenamefont {McKenzie}(1998)}]{McKenzie1998}%
	\BibitemOpen
	\bibfield  {author} {\bibinfo {author} {\bibfnamefont{R.}~\bibnamefont{H.}~\bibnamefont{McKenzie}},\ }\href {https://arxiv.org/abs/cond-mat/9802198}
	{\bibfield  {journal} {\bibinfo  {journal} {arXiv:cond-mat/9802198}}} [cond-mat.str-el] \BibitemShut {NoStop}%
	%
	\bibitem [{\citenamefont {Powell}\ and\ \citenamefont
		{McKenzie}(2006)}]{Powell2006}%
	\BibitemOpen
	\bibfield  {author} {\bibinfo {author} {\bibfnamefont {B.~J.}\ \bibnamefont
			{Powell}}\ and\ \bibinfo {author} {\bibfnamefont {R.~H.}\ \bibnamefont
			{McKenzie}},\ }\href
	{https://doi.org/10.1088{\%}2F0953-8984{\%}2F18{\%}2F45{\%}2Fr03} {\bibfield
		{journal} {\bibinfo  {journal} {J. Phys.: Condens. Matter}\ }\textbf
		{\bibinfo {volume} {18}},\ \bibinfo {pages} {R827} (\bibinfo {year}
		{2006})}\BibitemShut {NoStop}%
	\bibitem [{\citenamefont {Jacko}\ \emph {et~al.}(2020)\citenamefont {Jacko},
		\citenamefont {Kenny},\ and\ \citenamefont {Powell}}]{Jacko2019}%
	\BibitemOpen
	\bibfield  {author} {\bibinfo {author} {\bibfnamefont {A.~C.}\ \bibnamefont
			{Jacko}}, \bibinfo {author} {\bibfnamefont {E.~P.}\ \bibnamefont {Kenny}},\
		and\ \bibinfo {author} {\bibfnamefont {B.~J.}\ \bibnamefont {Powell}},\
	}\href {https://doi.org/10.1103/PhysRevB.101.125110} {\bibfield  {journal}
		{\bibinfo  {journal} {Phys. Rev. B}\ }\textbf {\bibinfo {volume} {101}},\
		\bibinfo {pages} {125110} (\bibinfo {year} {2020})}\BibitemShut {NoStop}%
	\bibitem [{\citenamefont {Janani}\ \emph
		{et~al.}(2014{\natexlab{b}})\citenamefont {Janani}, \citenamefont {Merino},
		\citenamefont {McCulloch},\ and\ \citenamefont {Powell}}]{Janani2014a}%
	\BibitemOpen
	\bibfield  {author} {\bibinfo {author} {\bibfnamefont {C.}~\bibnamefont
			{Janani}}, \bibinfo {author} {\bibfnamefont {J.}~\bibnamefont {Merino}},
		\bibinfo {author} {\bibfnamefont {I.~P.}\ \bibnamefont {McCulloch}},\ and\
		\bibinfo {author} {\bibfnamefont {B.~J.}\ \bibnamefont {Powell}},\ }\href
	{http://link.aps.org/doi/10.1103/PhysRevB.90.035120} {\bibfield  {journal}
		{\bibinfo  {journal} {Phys. Rev. B}\ }\textbf {\bibinfo {volume} {90}},\
		\bibinfo {pages} {35120} (\bibinfo {year} {2014}{\natexlab{b}})}\BibitemShut
	{NoStop}%
	\bibitem [{\citenamefont {Mattsson}\ \emph {et~al.}(1994)\citenamefont
		{Mattsson}, \citenamefont {Fr{\"{o}}jdh},\ and\ \citenamefont
		{Einarsson}}]{Mattsson1994}%
	\BibitemOpen
	\bibfield  {author} {\bibinfo {author} {\bibfnamefont {A.}~\bibnamefont
			{Mattsson}}, \bibinfo {author} {\bibfnamefont {P.}~\bibnamefont
			{Fr{\"{o}}jdh}},\ and\ \bibinfo {author} {\bibfnamefont {T.}~\bibnamefont
			{Einarsson}},\ }\href {https://link.aps.org/doi/10.1103/PhysRevB.49.3997}
	{\bibfield  {journal} {\bibinfo  {journal} {Phys. Rev. B}\ }\textbf {\bibinfo
			{volume} {49}},\ \bibinfo {pages} {3997} (\bibinfo {year}
		{1994})}\BibitemShut {NoStop}%
	\bibitem [{\citenamefont {Banerjee}\ \emph {et~al.}(2011)\citenamefont
		{Banerjee}, \citenamefont {Damle},\ and\ \citenamefont
		{Paramekanti}}]{Banerjee2011}%
	\BibitemOpen
	\bibfield  {author} {\bibinfo {author} {\bibfnamefont {A.}~\bibnamefont
			{Banerjee}}, \bibinfo {author} {\bibfnamefont {K.}~\bibnamefont {Damle}},\
		and\ \bibinfo {author} {\bibfnamefont {A.}~\bibnamefont {Paramekanti}},\
	}\href {https://link.aps.org/doi/10.1103/PhysRevB.83.134419} {\bibfield
		{journal} {\bibinfo  {journal} {Phys. Rev. B}\ }\textbf {\bibinfo {volume}
			{83}},\ \bibinfo {pages} {134419} (\bibinfo {year} {2011})}\BibitemShut
	{NoStop}%
	\bibitem [{\citenamefont {Pujari}\ \emph {et~al.}(2013)\citenamefont {Pujari},
		\citenamefont {Damle},\ and\ \citenamefont {Alet}}]{Pujari2013}%
	\BibitemOpen
	\bibfield  {author} {\bibinfo {author} {\bibfnamefont {S.}~\bibnamefont
			{Pujari}}, \bibinfo {author} {\bibfnamefont {K.}~\bibnamefont {Damle}},\ and\
		\bibinfo {author} {\bibfnamefont {F.}~\bibnamefont {Alet}},\ }\href
	{https://link.aps.org/doi/10.1103/PhysRevLett.111.087203} {\bibfield
		{journal} {\bibinfo  {journal} {Phys. Rev. Lett.}\ }\textbf {\bibinfo
			{volume} {111}},\ \bibinfo {pages} {087203} (\bibinfo {year}
		{2013})}\BibitemShut {NoStop}%
	\bibitem [{\citenamefont {Block}\ \emph {et~al.}(2013)\citenamefont {Block},
		\citenamefont {Melko},\ and\ \citenamefont {Kaul}}]{Block2013}%
	\BibitemOpen
	\bibfield  {author} {\bibinfo {author} {\bibfnamefont {M.~S.}\ \bibnamefont
			{Block}}, \bibinfo {author} {\bibfnamefont {R.~G.}\ \bibnamefont {Melko}},\
		and\ \bibinfo {author} {\bibfnamefont {R.~K.}\ \bibnamefont {Kaul}},\ }\href
	{https://link.aps.org/doi/10.1103/PhysRevLett.111.137202} {\bibfield
		{journal} {\bibinfo  {journal} {Phys. Rev. Lett.}\ }\textbf {\bibinfo
			{volume} {111}},\ \bibinfo {pages} {137202} (\bibinfo {year}
		{2013})}\BibitemShut {NoStop}%
	\bibitem [{\citenamefont {Harada}\ \emph {et~al.}(2013)\citenamefont {Harada},
		\citenamefont {Suzuki}, \citenamefont {Okubo}, \citenamefont {Matsuo},
		\citenamefont {Lou}, \citenamefont {Watanabe}, \citenamefont {Todo},\ and\
		\citenamefont {Kawashima}}]{Harada2013}%
	\BibitemOpen
	\bibfield  {author} {\bibinfo {author} {\bibfnamefont {K.}~\bibnamefont
			{Harada}}, \bibinfo {author} {\bibfnamefont {T.}~\bibnamefont {Suzuki}},
		\bibinfo {author} {\bibfnamefont {T.}~\bibnamefont {Okubo}}, \bibinfo
		{author} {\bibfnamefont {H.}~\bibnamefont {Matsuo}}, \bibinfo {author}
		{\bibfnamefont {J.}~\bibnamefont {Lou}}, \bibinfo {author} {\bibfnamefont
			{H.}~\bibnamefont {Watanabe}}, \bibinfo {author} {\bibfnamefont
			{S.}~\bibnamefont {Todo}},\ and\ \bibinfo {author} {\bibfnamefont
			{N.}~\bibnamefont {Kawashima}},\ }\href
	{https://link.aps.org/doi/10.1103/PhysRevB.88.220408} {\bibfield  {journal}
		{\bibinfo  {journal} {Phys. Rev. B}\ }\textbf {\bibinfo {volume} {88}},\
		\bibinfo {pages} {220408(R)} (\bibinfo {year} {2013})}\BibitemShut {NoStop}%
	\bibitem [{\citenamefont {Gong}\ \emph {et~al.}(2013)\citenamefont {Gong},
		\citenamefont {Sheng}, \citenamefont {Motrunich},\ and\ \citenamefont
		{Fisher}}]{Gong2013}%
	\BibitemOpen
	\bibfield  {author} {\bibinfo {author} {\bibfnamefont {S.-S.}\ \bibnamefont
			{Gong}}, \bibinfo {author} {\bibfnamefont {D.~N.}\ \bibnamefont {Sheng}},
		\bibinfo {author} {\bibfnamefont {O.~I.}\ \bibnamefont {Motrunich}},\ and\
		\bibinfo {author} {\bibfnamefont {M.~P.~A.}\ \bibnamefont {Fisher}},\ }\href
	{https://link.aps.org/doi/10.1103/PhysRevB.88.165138} {\bibfield  {journal}
		{\bibinfo  {journal} {Phys. Rev. B}\ }\textbf {\bibinfo {volume} {88}},\
		\bibinfo {pages} {165138} (\bibinfo {year} {2013})}\BibitemShut {NoStop}%
	\bibitem [{\citenamefont {Yu}\ \emph {et~al.}(2014)\citenamefont {Yu},
		\citenamefont {Liu}, \citenamefont {Li},\ and\ \citenamefont {Zou}}]{Yu2014}%
	\BibitemOpen
	\bibfield  {author} {\bibinfo {author} {\bibfnamefont {X.-L.}\ \bibnamefont
			{Yu}}, \bibinfo {author} {\bibfnamefont {D.-Y.}\ \bibnamefont {Liu}},
		\bibinfo {author} {\bibfnamefont {P.}~\bibnamefont {Li}},\ and\ \bibinfo
		{author} {\bibfnamefont {L.-J.}\ \bibnamefont {Zou}},\ }\href
	{http://www.sciencedirect.com/science/article/pii/S1386947713004542}
	{\bibfield  {journal} {\bibinfo  {journal} {Phys. E}\ }\textbf {\bibinfo
			{volume} {59}},\ \bibinfo {pages} {41} (\bibinfo {year} {2014})}\BibitemShut
	{NoStop}%
	\bibitem [{\citenamefont {Bishop}\ \emph {et~al.}(2015)\citenamefont {Bishop},
		\citenamefont {Li}, \citenamefont {G{\"{o}}tze}, \citenamefont {Richter},\
		and\ \citenamefont {Campbell}}]{Bishop2015}%
	\BibitemOpen
	\bibfield  {author} {\bibinfo {author} {\bibfnamefont {R.~F.}\ \bibnamefont
			{Bishop}}, \bibinfo {author} {\bibfnamefont {P.~H.~Y.}\ \bibnamefont {Li}},
		\bibinfo {author} {\bibfnamefont {O.}~\bibnamefont {G{\"{o}}tze}}, \bibinfo
		{author} {\bibfnamefont {J.}~\bibnamefont {Richter}},\ and\ \bibinfo {author}
		{\bibfnamefont {C.~E.}\ \bibnamefont {Campbell}},\ }\href
	{https://link.aps.org/doi/10.1103/PhysRevB.92.224434} {\bibfield  {journal}
		{\bibinfo  {journal} {Phys. Rev. B}\ }\textbf {\bibinfo {volume} {92}},\
		\bibinfo {pages} {224434} (\bibinfo {year} {2015})}\BibitemShut {NoStop}%
	\bibitem [{\citenamefont {Merino}\ \emph {et~al.}(2016)\citenamefont {Merino},
		\citenamefont {Jacko}, \citenamefont {Khosla},\ and\ \citenamefont
		{Powell}}]{Merino2016}%
	\BibitemOpen
	\bibfield  {author} {\bibinfo {author} {\bibfnamefont {J.}~\bibnamefont
			{Merino}}, \bibinfo {author} {\bibfnamefont {A.~C.}\ \bibnamefont {Jacko}},
		\bibinfo {author} {\bibfnamefont {A.~L.}\ \bibnamefont {Khosla}},\ and\
		\bibinfo {author} {\bibfnamefont {B.~J.}\ \bibnamefont {Powell}},\ }\href
	{https://doi.org/10.1103/PhysRevB.94.205109} {\bibfield  {journal} {\bibinfo
			{journal} {Phys. Rev. B}\ }\textbf {\bibinfo {volume} {94}},\ \bibinfo
		{pages} {205109} (\bibinfo {year} {2016})}\BibitemShut {NoStop}%
	\bibitem [{\citenamefont {Merino}\ \emph {et~al.}(2018)\citenamefont {Merino},
		\citenamefont {Jacko}, \citenamefont {Khosla}, \citenamefont {Ralko},\ and\
		\citenamefont {Powell}}]{Merino2018}%
	\BibitemOpen
	\bibfield  {author} {\bibinfo {author} {\bibfnamefont {J.}~\bibnamefont
			{Merino}}, \bibinfo {author} {\bibfnamefont {A.~C.}\ \bibnamefont {Jacko}},
		\bibinfo {author} {\bibfnamefont {A.~L.}\ \bibnamefont {Khosla}}, \bibinfo
		{author} {\bibfnamefont {A.}~\bibnamefont {Ralko}},\ and\ \bibinfo {author}
		{\bibfnamefont {B.~J.}\ \bibnamefont {Powell}},\ }\href
	{https://doi.org/10.1063/1.5041341} {\bibfield  {journal} {\bibinfo
			{journal} {AIP Adv.}\ }\textbf {\bibinfo {volume} {8}},\ \bibinfo {pages}
		{101430} (\bibinfo {year} {2018})}\BibitemShut {NoStop}%
	\bibitem [{\citenamefont {Merino}\ and\ \citenamefont
		{Ralko}(2018)}]{Merino2018b}%
	\BibitemOpen
	\bibfield  {author} {\bibinfo {author} {\bibfnamefont {J.}~\bibnamefont
			{Merino}}\ and\ \bibinfo {author} {\bibfnamefont {A.}~\bibnamefont {Ralko}},\
	}\href {https://link.aps.org/doi/10.1103/PhysRevB.97.205112} {\bibfield
		{journal} {\bibinfo  {journal} {Phys. Rev. B}\ }\textbf {\bibinfo {volume}
			{97}},\ \bibinfo {pages} {205112} (\bibinfo {year} {2018})}\BibitemShut
	{NoStop}%
	\bibitem [{\citenamefont {Gong}\ \emph {et~al.}(2015)\citenamefont {Gong},
		\citenamefont {Zhu},\ and\ \citenamefont {Sheng}}]{Gong2015}%
	\BibitemOpen
	\bibfield  {author} {\bibinfo {author} {\bibfnamefont {S.-S.}\ \bibnamefont
			{Gong}}, \bibinfo {author} {\bibfnamefont {W.}~\bibnamefont {Zhu}},\ and\
		\bibinfo {author} {\bibfnamefont {D.~N.}\ \bibnamefont {Sheng}},\ }\href
	{https://link.aps.org/doi/10.1103/PhysRevB.92.195110} {\bibfield  {journal}
		{\bibinfo  {journal} {Phys. Rev. B}\ }\textbf {\bibinfo {volume} {92}},\
		\bibinfo {pages} {195110} (\bibinfo {year} {2015})}\BibitemShut {NoStop}%
	\bibitem [{\citenamefont {Li}\ \emph {et~al.}(2016)\citenamefont {Li},
		\citenamefont {Bishop},\ and\ \citenamefont {Campbell}}]{Campbell2016}%
	\BibitemOpen
	\bibfield  {author} {\bibinfo {author} {\bibfnamefont {P.~H.~Y.}\
			\bibnamefont {Li}}, \bibinfo {author} {\bibfnamefont {R.~F.}\ \bibnamefont
			{Bishop}},\ and\ \bibinfo {author} {\bibfnamefont {C.~E.}\ \bibnamefont
			{Campbell}},\ }\href {http://stacks.iop.org/1742-6596/702/i=1/a=012001}
	{\bibfield  {journal} {\bibinfo  {journal} {J. Phys.: Conf. Ser.}\ }\textbf
		{\bibinfo {volume} {702}},\ \bibinfo {pages} {12001} (\bibinfo {year}
		{2016})}\BibitemShut {NoStop}%
	\bibitem [{\citenamefont {Mielke}(1991)}]{Mielke1991}%
	\BibitemOpen
	\bibfield  {author} {\bibinfo {author} {\bibfnamefont {A.}~\bibnamefont
			{Mielke}},\ }\href {http://stacks.iop.org/0305-4470/24/i=14/a=018} {\bibfield
		{journal} {\bibinfo  {journal} {J. Phys. A: Math. Gen.}\ }\textbf {\bibinfo
			{volume} {24}},\ \bibinfo {pages} {3311} (\bibinfo {year}
		{1991})}\BibitemShut {NoStop}%
	\bibitem [{\citenamefont {Mielke}(1992)}]{Mielke1992}%
	\BibitemOpen
	\bibfield  {author} {\bibinfo {author} {\bibfnamefont {A.}~\bibnamefont
			{Mielke}},\ }\href {http://stacks.iop.org/0305-4470/25/i=16/a=011} {\bibfield
		{journal} {\bibinfo  {journal} {J. Phys. A: Math. Gen.}\ }\textbf {\bibinfo
			{volume} {25}},\ \bibinfo {pages} {4335} (\bibinfo {year}
		{1992})}\BibitemShut {NoStop}%
	\bibitem [{\citenamefont {Tasaki}(1992)}]{Tasaki1992}%
	\BibitemOpen
	\bibfield  {author} {\bibinfo {author} {\bibfnamefont {H.}~\bibnamefont
			{Tasaki}},\ }\href {https://link.aps.org/doi/10.1103/PhysRevLett.69.1608}
	{\bibfield  {journal} {\bibinfo  {journal} {Phys. Rev. Lett.}\ }\textbf
		{\bibinfo {volume} {69}},\ \bibinfo {pages} {1608} (\bibinfo {year}
		{1992})}\BibitemShut {NoStop}%
	\bibitem [{\citenamefont {Stoner}(1951)}]{Stoner1951}%
	\BibitemOpen
	\bibfield  {author} {\bibinfo {author} {\bibfnamefont {E.~C.}\ \bibnamefont
			{Stoner}},\ }\href@noop {} {\bibfield  {journal} {\bibinfo  {journal} {J.
				Phys. Rad.}\ }\textbf {\bibinfo {volume} {12}},\ \bibinfo {pages} {372}
		(\bibinfo {year} {1951})}\BibitemShut {NoStop}%
	\bibitem [{\citenamefont {Powell}\ and\ \citenamefont
		{McKenzie}(2005)}]{Powell2005}%
	\BibitemOpen
	\bibfield  {author} {\bibinfo {author} {\bibfnamefont {B.~J.}\ \bibnamefont
			{Powell}}\ and\ \bibinfo {author} {\bibfnamefont {R.~H.}\ \bibnamefont
			{McKenzie}},\ }\href {https://doi.org/10.1103/PhysRevLett.94.047004}
	{\bibfield  {journal} {\bibinfo  {journal} {Phys. Rev. Lett.}\ }\textbf
		{\bibinfo {volume} {94}},\ \bibinfo {pages} {047004} (\bibinfo {year}
		{2005})}\BibitemShut {NoStop}%
	\bibitem [{\citenamefont {Powell}\ and\ \citenamefont
		{McKenzie}(2007)}]{Powell2007}%
	\BibitemOpen
	\bibfield  {author} {\bibinfo {author} {\bibfnamefont {B.~J.}\ \bibnamefont
			{Powell}}\ and\ \bibinfo {author} {\bibfnamefont {R.~H.}\ \bibnamefont
			{McKenzie}},\ }\href {https://doi.org/10.1103/PhysRevLett.98.027005}
	{\bibfield  {journal} {\bibinfo  {journal} {Phys. Rev. Lett.}\ }\textbf
		{\bibinfo {volume} {98}},\ \bibinfo {pages} {027005} (\bibinfo {year}
		{2007})}\BibitemShut {NoStop}%
	\bibitem [{\citenamefont {Hoshino}\ and\ \citenamefont
		{Werner}(2015)}]{Hoshino2015}%
	\BibitemOpen
	\bibfield  {author} {\bibinfo {author} {\bibfnamefont {S.}~\bibnamefont
			{Hoshino}}\ and\ \bibinfo {author} {\bibfnamefont {P.}~\bibnamefont
			{Werner}},\ }\href {https://link.aps.org/doi/10.1103/PhysRevLett.115.247001}
	{\bibfield  {journal} {\bibinfo  {journal} {Phys. Rev. Lett.}\ }\textbf
		{\bibinfo {volume} {115}},\ \bibinfo {pages} {247001} (\bibinfo {year}
		{2015})}\BibitemShut {NoStop}%
	\bibitem [{\citenamefont {Hoshino}\ and\ \citenamefont
		{Werner}(2016)}]{Hoshino2016}%
	\BibitemOpen
	\bibfield  {author} {\bibinfo {author} {\bibfnamefont {S.}~\bibnamefont
			{Hoshino}}\ and\ \bibinfo {author} {\bibfnamefont {P.}~\bibnamefont
			{Werner}},\ }\href {https://link.aps.org/doi/10.1103/PhysRevB.93.155161}
	{\bibfield  {journal} {\bibinfo  {journal} {Phys. Rev. B}\ }\textbf {\bibinfo
			{volume} {93}},\ \bibinfo {pages} {155161} (\bibinfo {year}
		{2016})}\BibitemShut {NoStop}%
	%
	\bibitem{Merino2020}
	\BibitemOpen
	\bibfield  {author} {\bibinfo {author} {\bibfnamefont{J.}~\bibnamefont{Merino}},\ \bibinfo {author} {\bibfnamefont {M.}~\bibnamefont{F.}~\bibnamefont{L\'{o}pez}},\ and\ \bibinfo {author} {\bibnamefont{B.}~\bibnamefont{J.}~\bibnamefont{Powell}},\ }\href {https://arxiv.org/abs/2012.13211}
	{\bibfield  {journal} {\bibinfo  {journal} {arXiv:2012.13211}}} [cond-mat.supr-con] \BibitemShut {NoStop}%
\end{thebibliography}

\begin{thebibliography}{11}%
	\makeatletter
	\providecommand \@ifxundefined [1]{%
		\@ifx{#1\undefined}
	}%
	\providecommand \@ifnum [1]{%
		\ifnum #1\expandafter \@firstoftwo
		\else \expandafter \@secondoftwo
		\fi
	}%
	\providecommand \@ifx [1]{%
		\ifx #1\expandafter \@firstoftwo
		\else \expandafter \@secondoftwo
		\fi
	}%
	\providecommand \natexlab [1]{#1}%
	\providecommand \enquote  [1]{``#1''}%
	\providecommand \bibnamefont  [1]{#1}%
	\providecommand \bibfnamefont [1]{#1}%
	\providecommand \citenamefont [1]{#1}%
	\providecommand \href@noop [0]{\@secondoftwo}%
	\providecommand \href [0]{\begingroup \@sanitize@url \@href}%
	\providecommand \@href[1]{\@@startlink{#1}\@@href}%
	\providecommand \@@href[1]{\endgroup#1\@@endlink}%
	\providecommand \@sanitize@url [0]{\catcode `\\12\catcode `\$12\catcode
		`\&12\catcode `\#12\catcode `\^12\catcode `\_12\catcode `\%12\relax}%
	\providecommand \@@startlink[1]{}%
	\providecommand \@@endlink[0]{}%
	\providecommand \url  [0]{\begingroup\@sanitize@url \@url }%
	\providecommand \@url [1]{\endgroup\@href {#1}{\urlprefix }}%
	\providecommand \urlprefix  [0]{URL }%
	\providecommand \Eprint [0]{\href }%
	\providecommand \doibase [0]{https://doi.org/}%
	\providecommand \selectlanguage [0]{\@gobble}%
	\providecommand \bibinfo  [0]{\@secondoftwo}%
	\providecommand \bibfield  [0]{\@secondoftwo}%
	\providecommand \translation [1]{[#1]}%
	\providecommand \BibitemOpen [0]{}%
	\providecommand \bibitemStop [0]{}%
	\providecommand \bibitemNoStop [0]{.\EOS\space}%
	\providecommand \EOS [0]{\spacefactor3000\relax}%
	\providecommand \BibitemShut  [1]{\csname bibitem#1\endcsname}%
	\let\auto@bib@innerbib\@empty
	%</preamble>
	\bibitem [{\citenamefont {Lechermann}\ \emph {et~al.}(2007)\citenamefont
		{Lechermann}, \citenamefont {Georges}, \citenamefont {Kotliar},\ and\
		\citenamefont {Parcollet}}]{SLechermann2007}%
	\BibitemOpen
	\bibfield  {author} {\bibinfo {author} {\bibfnamefont {F.}~\bibnamefont
			{Lechermann}}, \bibinfo {author} {\bibfnamefont {A.}~\bibnamefont {Georges}},
		\bibinfo {author} {\bibfnamefont {G.}~\bibnamefont {Kotliar}},\ and\ \bibinfo
		{author} {\bibfnamefont {O.}~\bibnamefont {Parcollet}},\ }\href
	{http://link.aps.org/doi/10.1103/PhysRevB.76.155102} {\bibfield  {journal}
		{\bibinfo  {journal} {Phys. Rev. B}\ }\textbf {\bibinfo {volume} {76}},\
		\bibinfo {pages} {155102} (\bibinfo {year} {2007})}\BibitemShut {NoStop}%
	\bibitem [{\citenamefont {Isidori}\ and\ \citenamefont
		{Capone}(2009)}]{SIsidori2009}%
	\BibitemOpen
	\bibfield  {author} {\bibinfo {author} {\bibfnamefont {A.}~\bibnamefont
			{Isidori}}\ and\ \bibinfo {author} {\bibfnamefont {M.}~\bibnamefont
			{Capone}},\ }\href {https://doi.org/10.1103/PhysRevB.80.115120} {\bibfield
		{journal} {\bibinfo  {journal} {Phys. Rev. B}\ }\textbf {\bibinfo {volume}
			{80}},\ \bibinfo {pages} {115120} (\bibinfo {year} {2009})}\BibitemShut
	{NoStop}%
	\bibitem [{\citenamefont {Lanat{\`{a}}}\ \emph {et~al.}(2015)\citenamefont
		{Lanat{\`{a}}}, \citenamefont {Yao}, \citenamefont {Wang}, \citenamefont
		{Ho},\ and\ \citenamefont {Kotliar}}]{SLanata2015}%
	\BibitemOpen
	\bibfield  {author} {\bibinfo {author} {\bibfnamefont {N.}~\bibnamefont
			{Lanat{\`{a}}}}, \bibinfo {author} {\bibfnamefont {Y.}~\bibnamefont {Yao}},
		\bibinfo {author} {\bibfnamefont {C.-Z.}\ \bibnamefont {Wang}}, \bibinfo
		{author} {\bibfnamefont {K.-M.}\ \bibnamefont {Ho}},\ and\ \bibinfo {author}
		{\bibfnamefont {G.}~\bibnamefont {Kotliar}},\ }\href
	{http://link.aps.org/doi/10.1103/PhysRevX.5.011008} {\bibfield  {journal}
		{\bibinfo  {journal} {Phys. Rev. X}\ }\textbf {\bibinfo {volume} {5}},\
		\bibinfo {pages} {11008} (\bibinfo {year} {2015})}\BibitemShut {NoStop}%
	\bibitem [{\citenamefont {Lanat{\`{a}}}\ \emph {et~al.}(2017)\citenamefont
		{Lanat{\`{a}}}, \citenamefont {Yao}, \citenamefont {Deng}, \citenamefont
		{Dobrosavljevi{\'{c}}},\ and\ \citenamefont {Kotliar}}]{SLanata2017}%
	\BibitemOpen
	\bibfield  {author} {\bibinfo {author} {\bibfnamefont {N.}~\bibnamefont
			{Lanat{\`{a}}}}, \bibinfo {author} {\bibfnamefont {Y.}~\bibnamefont {Yao}},
		\bibinfo {author} {\bibfnamefont {X.}~\bibnamefont {Deng}}, \bibinfo {author}
		{\bibfnamefont {V.}~\bibnamefont {Dobrosavljevi{\'{c}}}},\ and\ \bibinfo
		{author} {\bibfnamefont {G.}~\bibnamefont {Kotliar}},\ }\href
	{https://link.aps.org/doi/10.1103/PhysRevLett.118.126401} {\bibfield
		{journal} {\bibinfo  {journal} {Phys. Rev. Lett.}\ }\textbf {\bibinfo
			{volume} {118}},\ \bibinfo {pages} {126401} (\bibinfo {year}
		{2017})}\BibitemShut {NoStop}%
	\bibitem [{\citenamefont {Kotliar}\ and\ \citenamefont
		{Ruckenstein}(1986)}]{SKotliar1986}%
	\BibitemOpen
	\bibfield  {author} {\bibinfo {author} {\bibfnamefont {G.}~\bibnamefont
			{Kotliar}}\ and\ \bibinfo {author} {\bibfnamefont {A.~E.}\ \bibnamefont
			{Ruckenstein}},\ }\href {http://link.aps.org/doi/10.1103/PhysRevLett.57.1362}
	{\bibfield  {journal} {\bibinfo  {journal} {Phys. Rev. Lett.}\ }\textbf
		{\bibinfo {volume} {57}},\ \bibinfo {pages} {1362} (\bibinfo {year}
		{1986})}\BibitemShut {NoStop}%
	\bibitem [{\citenamefont {B\"unemann}\ \emph {et~al.}(2003)\citenamefont
		{B\"unemann}, \citenamefont {Gebhard},\ and\ \citenamefont
		{Thul}}]{SBunemann2003}%
	\BibitemOpen
	\bibfield  {author} {\bibinfo {author} {\bibfnamefont {J.}~\bibnamefont
			{B\"unemann}}, \bibinfo {author} {\bibfnamefont {F.}~\bibnamefont {Gebhard}},\
		and\ \bibinfo {author} {\bibfnamefont {R.}~\bibnamefont {Thul}},\ }\href
	{https://doi.org/10.1103/PhysRevB.67.075103} {\bibfield  {journal} {\bibinfo
			{journal} {Phys. Rev. B}\ }\textbf {\bibinfo {volume} {67}},\ \bibinfo
		{pages} {75103} (\bibinfo {year} {2003})}\BibitemShut {NoStop}%
	\bibitem [{\citenamefont {Lanat{\`{a}}}\ \emph {et~al.}(2017)\citenamefont
		{Lanat{\`{a}}}, \citenamefont {Lee}, \citenamefont {Yao},\ and\ \citenamefont
		{Dobrosavljevi{\'{c}}}}]{SLanata2017b}%
	\BibitemOpen
	\bibfield  {author} {\bibinfo {author} {\bibfnamefont {N.}~\bibnamefont
			{Lanat{\`{a}}}}, \bibinfo {author} {\bibfnamefont {T.-H.}\ \bibnamefont {Lee}},
		\bibinfo {author} {\bibfnamefont {Y.-X.}\ \bibnamefont {Yao}},\ and\ \bibinfo
		{author} {\bibfnamefont {V.}~\bibnamefont {Dobrosavljevi{\'{c}}}},\ }\href
	{https://doi.org/10.1103/PhysRevB.96.195126} {\bibfield  {journal} {\bibinfo
			{journal} {Phys. Rev. B}\ }\textbf {\bibinfo {volume} {96}},\ \bibinfo
		{pages} {195126} (\bibinfo {year} {2017})}\BibitemShut {NoStop}%
	\bibitem [{\citenamefont {Parcollet}\ \emph {et~al.}(2015)\citenamefont
		{Parcollet}, \citenamefont {Ferrero}, \citenamefont {Ayral}, \citenamefont
		{Hafermann}, \citenamefont {Krivenko}, \citenamefont {Messio},\ and\
		\citenamefont {Seth}}]{SParcollet2015}%
	\BibitemOpen
	\bibfield  {author} {\bibinfo {author} {\bibfnamefont {O.}~\bibnamefont
			{Parcollet}}, \bibinfo {author} {\bibfnamefont {M.}~\bibnamefont {Ferrero}},
		\bibinfo {author} {\bibfnamefont {T.}~\bibnamefont {Ayral}}, \bibinfo
		{author} {\bibfnamefont {H.}~\bibnamefont {Hafermann}}, \bibinfo {author}
		{\bibfnamefont {I.}~\bibnamefont {Krivenko}}, \bibinfo {author}
		{\bibfnamefont {L.}~\bibnamefont {Messio}},\ and\ \bibinfo {author}
		{\bibfnamefont {P.}~\bibnamefont {Seth}},\ }\href
	{http://www.sciencedirect.com/science/article/pii/S0010465515001666}
	{\bibfield  {journal} {\bibinfo  {journal} {Comput. Phys. Commun.}\ }\textbf
		{\bibinfo {volume} {196}},\ \bibinfo {pages} {398} (\bibinfo {year}
		{2015})},\ \bibinfo {note} {version 1.4}\BibitemShut {NoStop}%
	\bibitem [{\citenamefont {Seth}\ \emph {et~al.}(2016)\citenamefont {Seth},
		\citenamefont {Krivenko}, \citenamefont {Ferrero},\ and\ \citenamefont
		{Parcollet}}]{SSeth2016}%
	\BibitemOpen
	\bibfield  {author} {\bibinfo {author} {\bibfnamefont {P.}~\bibnamefont
			{Seth}}, \bibinfo {author} {\bibfnamefont {I.}~\bibnamefont {Krivenko}},
		\bibinfo {author} {\bibfnamefont {M.}~\bibnamefont {Ferrero}},\ and\ \bibinfo
		{author} {\bibfnamefont {O.}~\bibnamefont {Parcollet}},\ }\href
	{http://www.sciencedirect.com/science/article/pii/S001046551500404X}
	{\bibfield  {journal} {\bibinfo  {journal} {Comput. Phys. Commun.}\ }\textbf
		{\bibinfo {volume} {200}},\ \bibinfo {pages} {274} (\bibinfo {year}
		{2016})}\BibitemShut {NoStop}%
	\bibitem [{\citenamefont {Bl{\"{o}}chl}\ \emph {et~al.}(1994)\citenamefont
		{Bl{\"{o}}chl}, \citenamefont {Jepsen},\ and\ \citenamefont
		{Andersen}}]{SBlochl1994}%
	\BibitemOpen
	\bibfield  {author} {\bibinfo {author} {\bibfnamefont {P.~E.}\ \bibnamefont
			{Bl{\"{o}}chl}}, \bibinfo {author} {\bibfnamefont {O.}~\bibnamefont
			{Jepsen}},\ and\ \bibinfo {author} {\bibfnamefont {O.~K.}\ \bibnamefont
			{Andersen}},\ }\href {http://link.aps.org/doi/10.1103/PhysRevB.49.16223}
	{\bibfield  {journal} {\bibinfo  {journal} {Phys. Rev. B}\ }\textbf {\bibinfo
			{volume} {49}},\ \bibinfo {pages} {16223} (\bibinfo {year}
		{1994})}\BibitemShut {NoStop}%
	\bibitem [{\citenamefont {Krivenko}(2020)}]{STRIQS-ARPACK}%
	\BibitemOpen
	\bibfield  {author} {\bibinfo {author} {\bibfnamefont {I.}~\bibnamefont
			{Krivenko}},\ }\href {https://doi.org/10.5281/zenodo.3930202} {\bibinfo
		{title} {{ezARPACK - a C++ ARPACK-NG wrapper compatible with multiple
				matrix/vector algebra libraries: Release 0.9}}} (\bibinfo {year}
	{2020})\BibitemShut {NoStop}%
\end{thebibliography}

%

\end{document}